\documentclass[prb,showpacs,preprintnumbers,amsfonts,amssymb,amsmath,floats,twocolumn,aps]{revtex4}

\usepackage{graphicx}
\usepackage{dcolumn}
\usepackage{nicefrac}
\usepackage{bm}
\usepackage{color}
\newcommand{\ff}[1]{{\boldsymbol #1}}

\newcommand{\ba}{\begin{eqnarray}}
\newcommand{\ea}{\end{eqnarray}}
\newcommand{\be}{\begin{equation}}
\newcommand{\ee}{\end{equation}}
\newcommand{\bi}{\begin{itemize}}
\newcommand{\ei}{\end{itemize}}

\begin{document} 
  
\title{Screening mechanisms in magnetic nanostructures}

\author{Andrej Schwabe, Mirek H\"ansel and Michael Potthoff}
\affiliation{I. Institut f\"ur Theoretische Physik, Universit\"at Hamburg, Jungiusstra\ss{}e 9, 20355 Hamburg, Germany}

\author{Andrew K. Mitchell}
\affiliation{Institute for Theoretical Physics, Utrecht University, Leuvenlaan 4, 3584 CE Utrecht, The Netherlands}

\begin{abstract}
A theoretical concept is presented for the screening of several magnetic moments locally exchange coupled to conduction electrons in a metallic nanostructure. 
We consider a quantum confined multi-impurity Kondo model which exhibits the competition between finite-size effects, RKKY interactions, and Kondo physics. 
In the limit of weak coupling, Kondo correlations are cut by the finite system size; perturbation theory can then be used to derive the low-energy effective model, which is of generalized central-spin form. 
The theory successfully predicts the degeneracy, total spin, and spin correlations of the ground state, and allows the number of screening channels to be identified. 
This is demonstrated for a two-impurity model on a finite one-dimensional ring. 
Density-matrix renormalization-group calculations confirm the physical picture at weak coupling. 
The non-trivial crossovers to RKKY and strong-coupling regimes are also studied. 
The numerical renormalization group, tailored to treat finite systems, is used to examine the crossover to the thermodynamic limit. 
\end{abstract} 
 

% 75.20.Hr	Local moment in compounds and alloys; Kondo effect, valence fluctuations, heavy fermions
% 75.75.-c 	Magnetic properties of nanostructures
% 71.70.Gm	Exchange interactions

\maketitle 

%--------------------------------------------------------------------------------------------------------------
\section{Introduction}
%--------------------------------------------------------------------------------------------------------------

Many of the fascinating properties of correlated electron systems derive from collective behavior. 
In systems with a magnetic impurity embedded in a host of conduction electrons, the Kondo effect \cite{Hew93} represents a prime example. 
It is caused by the antiferromagnetic exchange interaction between the impurity spin and the local spin of conduction electrons.
Below a characteristic temperature $T_{\rm K}$, the Kondo temperature, the magnetic moment of the impurity is collectively screened by a cloud of conduction electrons with spatial extension determined by $\xi_{\rm K} \sim v_{\rm F} / T_{\rm K}$, where $v_{\rm F}$ is the Fermi velocity. \cite{barzykin1996kondo,MBB11}
Kondo screening is a phenomenon that eludes a perturbative approach, as is expressed by the non-analytic dependence of $T_{\rm K}$ on the strength $J$ of the local exchange coupling: $\ln T_{\rm K} \sim - 1/J$. 
For weak $J$, the Kondo scale is exponentially small. 
This translates into a mesoscopically large Kondo cloud.

Hence, for a magnetic impurity embedded in a small nanostructure, 
\cite{RBT95,RBT97,BWD+05,KZC+06,OHCL00,BBM+10}
the concept of Kondo screening must be modified. 
\cite{TKvD99,SA02,CB02,LVK05,HKM06}
The universal aspects of Kondo physics are lost when $T_{\rm K}$ is of the same order of magnitude as the finite-size gap $\Delta$ at the Fermi energy of the conduction-electron system or, if viewed from a real-space perspective, when the Kondo cloud exceeds the size of the system. 
A Kondo scale in the regime of some tens of Kelvins roughly corresponds to a linear extension in the nanometer range.

If there are several magnetic impurities,
\cite{CTL+04,ZWL+10,NBK+11}
the situation gets complicated by the competition of Kondo screening with non-local magnetic correlations.
The latter are typically induced by an indirect exchange coupling which is mediated by the conduction electrons.
The effective Ruderman-Kittel-Kasuya-Yosida (RKKY) interaction \cite{RK54,Kas56,Yos57} can be derived perturbatively and scales as $J_{\rm RKKY} \sim J^{2}$. 
In the weak-$J$ regime this dominates over the exponentially small Kondo scale, while the Kondo effect sets in for sufficiently strong $J>J_{\rm D}$. \cite{Don77,mitchell2015multiple,allerdt2015kondo} 

This competition between Kondo screening and RKKY interaction is also qualitatively modified by quantum confinement: \cite{SGP12,GBA04,CA02,NHdA07} 
For coupling strengths weaker than $J_{\Delta}$, where $J_{\Delta}$ is defined by the condition $\Delta=T_{K}(J_{\Delta})$, the conventional Kondo effect is replaced by a ``finite-size Kondo effect'' where an impurity spin is screened by forming a singlet with a \emph{single} conduction electron occupying the fully delocalized one-particle state at the Fermi energy.
The energy to break this singlet scales linearly (rather than exponentially) with $J$ and therefore the finite-size Kondo effect dominates over the RKKY interaction for $J\to 0$.
Hence, with decreasing $J$ there is a reentrance of the Kondo effect. \cite{SGP12}

In this paper we focus on the conditions under which different magnetic moments (impurity spins) in a nanostructure can be individually screened. 
To investigate possible screening mechanisms on the linear-in-$J$ scale, we consider a generic multi-impurity Kondo model involving a conduction-electron system with a finite number of sites.
In the weak-coupling limit $J < J_{\Delta}$, the physics of this model is accessible by standard perturbation theory. 
In particular, we study the case where the electronic structure of the non-interacting conduction-electron system exhibits more than one single-particle state at the Fermi energy $\varepsilon_{\rm F}$ due to symmetries of the confinement geometry. 
There are several questions to be answered in this context, e.g.:
Which states at $\varepsilon_{\rm F}$ provide a ``channel'' for the screening of an impurity spin?
How many impurity spins can be screened for $J \to 0$?
In which way does the geometry and the electronic structure of the nanosystem affect the screening?

Gapless Kondo models ($\Delta = 0$) with higher spin quantum numbers or multiple screening channels \cite{NB80,AL93,RFC+09,PRS+07,mitchell2012two} have attracted considerable interest in the past since their low-energy properties cannot be captured by local Fermi-liquid theory. \cite{NP66,Noz74}
The present paper addresses the simpler but analogous question: in a nanosystem with several magnetic impurities and degenerate orbitals at the Fermi energy, when and how does an overall spin-singlet ground state arise? Are there analogs of underscreening and overscreening?

A highly interesting question concerns the crossover from the physics of the ``finite-size multi-channel Kondo effect'' to the physics of the gapless system, realized in the thermodynamic limit. 
This crossover should be visible when extending the nanostructure in size, corresponding to the limit $\Delta \to 0$ and $J_{\Delta} \to 0$, or when increasing $J$ and thereby approaching the regime $J > J_{\Delta}$ where finite-size effects no longer matter.
With increasing coupling strength, the individual screening on the linear-in-$J$ scale is expected to be replaced by non-local magnetic correlations due to the operation of the RKKY coupling at order $J^{2}$ until, on further increasing the coupling toward the strong-$J$ limit, the standard Kondo effect is again recovered.

To elucidate our ideas with concrete examples we consider the two-impurity Kondo model on a one-dimensional lattice in a ring geometry.  
This model not only shows individual screening of the two impurity spins for $J\to 0$, depending on details of the geometry and electronic structure, but also exhibits various crossovers as a function of $J$ and of the system size.
These are studied numerically by employing the density-matrix renormalization group (DMRG) \cite{Whi92,Sch11} as well as the numerical renormalization group (NRG). \cite{Wil75,nrg_rev} 
We use a matrix-product implementation and a folded-chain geometry to account for the periodic boundary conditions within the DMRG. 
Application of the NRG implementation to an impurity (or multi-impurity) system with a discrete bath requires a specific adaption of the theory and the numerical algorithm to perform the Wilson-chain mappings.

The paper is organized as follows:
After briefly introducing the model in Sec.\ \ref{sec:model}, the general perturbative theory is explained in Sec.\ \ref{sec:theory}. 
Results obtained with the DMRG and the NRG for the two-impurity Kondo model on a finite ring are discussed in Secs.\ \ref{sec:dmrg} and \ref{sec:nrg}, respectively. 
A summary and the conclusions are given in Sec.\ \ref{sec:con}.

%--------------------------------------------------------------------------------------------------------------
\section{Multi-impurity Kondo model}
\label{sec:model}
%--------------------------------------------------------------------------------------------------------------

Using standard notations, the multi-impurity Kondo model on a lattice of finite size is given by the following Hamiltonian:
\be
   H = H_{0} + H_{1} 
       = \sum_{ii' \sigma} t_{ii'} c^\dagger_{i\sigma} c_{i'\sigma}
       + J \sum_{r=1}^R \ff s_{i_r} \cdot \mathbf S_r
       \: .
\label{eq:ham}
\ee
The first part $H_{0}$ describes a system of $N$ non-interacting conduction electrons moving on a lattice of arbitrary dimension and geometry consisting of $L$ sites. 
There is a spin-degenerate orbital $| i, \sigma \rangle$ at each site $i$, and $\sigma=\uparrow, \downarrow$ is the spin projection. 
$c_{i\sigma}^{\dagger}$ and $c_{i\sigma}$ are the corresponding creators and annihilators.
Since we wish to address the physics of magnetic nanostructures, we take $L$ to be finite. 
Diagonalization of the spin-independent tight-binding hopping matrix $t_{ii'}$,
\be
  \sum_{ii'} U^{\dagger}_{kg;i} t_{ii'} U_{i'; k'g'} = \varepsilon_{k} \delta_{kk'} \delta_{gg'}  
  \: ,
\label{eq:umatrix}  
\ee
yields the set of one-particle energies $\varepsilon_{k}$ and one-particle eigenstates $| k, g, \sigma \rangle$, where the index $g=1,...,G(k)$ accounts for the possible degeneracy of $\varepsilon_{k}$ arising, e.g., due to spatial symmetries of the nanostructure. 
The unitary matrix $\ff U$ has elements $U_{i; kg} = \langle i , \sigma | k,g,\sigma \rangle$ and is spin-independent. 
Three-dimensional systems with linear extension in the range of $\sim 10$~nm correspond to a level spacing $\Delta \varepsilon_{k} = \varepsilon_{k+1} - \varepsilon_{k}$ of the order of some tens of Kelvins.

The second part $H_{1}$ describes $R$ magnetic impurities, modeled as spins with spin quantum number $S=1/2$.
Each impurity is coupled by a local antiferromagnetic exchange $J>0$ to the conduction-electron system. 
The $r$-th spin $\ff S_{r}$ interacts with the local conduction-electron spin density $\ff s_{i_{r}} = \tfrac{1}{2} \sum_{\sigma\sigma'} c^\dagger_{i_{r}\sigma} \ff \sigma_{\sigma\sigma'} c_{i_{r}\sigma'}$ at site $i_{r}$.
Here, $\ff \sigma$ denotes the vector of Pauli matrices.

The physics of this model in the thermodynamic limit with $L\to \infty$ is highly complex and characterized by a subtle competition \cite{Don77,mitchell2015multiple,allerdt2015kondo} between Kondo screening on a scale given by the Kondo \cite{Hew93} temperature $T_{\rm K}$ and the indirect RKKY \cite{RK54,Kas56,Yos57} magnetic coupling $J_{\rm RKKY}$.
While the RKKY coupling is a perturbative concept and obtained with second-order-in-$J$ perturbation theory, the Kondo effect is non-perturbative with $\ln T_{\rm K} \sim -1/J$. 
Because the Kondo temperature is exponentially small, the RKKY coupling $J_{\rm RKKY} \sim J^{2}$ dominates in the weak-$J$ regime.

As was pointed out in Ref.\ \onlinecite{SGP12}, this picture may change qualitatively for systems of finite size $L$ when $T_{\rm K} \sim \Delta$, where $\Delta$ is the finite-size gap at the Fermi energy. 
In this case, the logarithmic Kondo correlations are cut by the finite system size, \cite{TKvD99} and the universal Kondo effect is replaced by a ``finite-size Kondo effect''. The latter is characterized by a linear-in-$J$ energy scale, whose precise value depends on details of the geometry and electronic structure, and which dominates over the RKKY interaction as $J\to 0$.
Since the finite-size gap $\Delta$ regularizes the problem, the ``finite-size Kondo effect'' should be well accessible to standard perturbation theory for coupling strengths $J < J_{\Delta}$, where $J_{\Delta}$ is given by the condition $\Delta=T_{K}(J_{\Delta})$. 
Since $\ln T_{K} \sim -1/J$ for the infinite system, \cite{Hew93} we have $J\lesssim -1/\ln \Delta $.

%--------------------------------------------------------------------------------------------------------------
\section{Effective low-energy theory}
\label{sec:theory}
%--------------------------------------------------------------------------------------------------------------

Let $| \rm FS, \gamma \rangle$, with $\gamma=1,...,\Gamma$, be the ($\Gamma$-fold degenerate) $N$-electron ground state of $H_{0}$ (the Fermi sea) which is obtained by occupying all one-particle levels below the Fermi energy, $\varepsilon_{k} \le \varepsilon_{\rm F}$.

One can distinguish between two situations: 
In the ``off-resonance'' case, the ground state is non-degenerate, $\Gamma=1$, and the orbitals $|k_{\rm F}, g, \sigma\rangle$ with $g=1,...,G(k_{\rm F})$ at $\varepsilon_{\rm F} = \varepsilon_{k_{\rm F}}$ are fully occupied.
Note, that this requires an even number of electrons in the nanostructure.
In this off-resonance case, the interaction $H_{1}$ can be treated as a perturbation to recover the effective RKKY model, 
\be
   H_{\rm RKKY} = \frac12 \sum_{rr'} J_{\rm RKKY, rr'} \ff S_{r} \cdot \ff S_{r'}
   \: 
\label{eq:rkky}
\ee
where $J_{\rm RKKY, rr'}$ is given by $J^{2}$ times the non-local ($r\ne r'$) static spin susceptibility of the non-interacting conduction-electron system.
In particular, there is no linear-in-$J$ Kondo effect in this case. 

In the following, we will concentrate on the ``on-resonance'' case where $\Gamma>1$, i.e., where the one-particle orbitals with energy $\varepsilon_{\rm F}$ are incompletely occupied. 
Note in particular that this is necessarily the case when the nanostructure hosts an odd number of electrons $N$.
Employing degenerate perturbation theory \cite{EFG+05} up to linear order in $J$, the low-energy physics is captured by an effective model
\be
   H_{\rm eff} = P_0 H_1 P_0 
   \: ,
\label{eq:effmodel}   
\ee
where $P_0 = \sum_{\gamma=1}^{\Gamma} | \rm FS,\gamma \rangle \langle \rm FS, \gamma |$ is a projector onto the space of ground states of the unperturbed ($J=0$) Hamiltonian. 
To compute $P_{0} \ff s_{i_{r}} P_{0}$, we first consider the unitary transformation 
\be
  c^{\dagger}_{kg\sigma} = \sum_{i} U_{i; kg} c^{\dagger}_{i\sigma}
  \: ,
\ee
which gives 
\be
\ff s_{i_{r}} 
=
\frac12
\sum_{kk',gg',\sigma\sigma'} 
U_{kg; i_{r}}^{\dagger} c^\dagger_{kg\sigma} 
\ff \sigma_{\sigma\sigma'} 
c_{k'g'\sigma'} 
U_{i_{r} ; k'g'}
\: .
\ee
Since the Pauli matrices are traceless, terms with $k,k' \ne k_{\rm F}$ do not contribute to $P_{0} \ff s_{i_{r}} P_{0}$:
\be
P_{0} 
\ff s_{i_{r}} 
P_{0}
=
\frac12
\sum_{gg',\sigma\sigma'} 
U_{k_{\rm F}g; i_{r}}^{\dagger} c^\dagger_{k_{\rm F}g\sigma} 
\ff \sigma_{\sigma\sigma'} 
c_{k_{\rm F}g'\sigma'} 
U_{i_{r} ; k_{\rm F}g'}
P_{0}
\: .
\ee
The effective low-energy model Eq.\ (\ref{eq:effmodel}) can be written as a spin-only model. 
To this end we introduce another set of {\em site-dependent} unitary transformations: 
\be
c^\dagger_{k_F\alpha\sigma}(i_r)
=
\sum_{g}
V_{g\alpha}(i_{r})
c^\dagger_{k_Fg\sigma}
\: .
\label{eq:str}
\ee
In general, the unitary $G(k_{\rm F}) \times G(k_{\rm F})$ matrices $\ff V(i_{r})$ are different for each site $i_{r}$ (and do not commute); the creator $c^\dagger_{k_F\alpha\sigma}(i_r)$ is therefore also dependent on $i_{r}$. 
$\ff V(i_{r})$ are the transformations that diagonalize the dyadic products 
$u_{gg'}(i_{r}) \equiv U^{\dagger}_{k_{\rm F} g; i_{r}} U_{i_r;k_{\rm F}g'}$ for each $i_{r}$ with $r = 1 ,..., R$:
\be
\sum_{gg'} 
V_{\alpha g}^{\dagger}(i_{r})
U^{\dagger}_{k_{\rm F} g; i_{r}} U_{i_r;k_{\rm F}g'}
V_{g'\alpha'}(i_{r})
=
x_{\alpha}(i_{r}) 
\delta_{\alpha\alpha'}
\: .
\ee
Trivially, this construction implies that there is a single non-zero eigenvalue,  $x_{\alpha=1}(i_{r}) = \sum_{g} |U_{i_r;k_{\rm F}g}|^{2}$. 
The corresponding eigenvector has the elements $V_{g\alpha=1}(i_{r}) = U^{\dagger}_{k_{\rm F}g; i_{r}} / 
\sqrt{\sum_{g} |U_{i_r;k_{\rm F}g}|^{2}}$.
Using this, and collecting the results, we find
\be
P_{0} 
\ff s_{i_{r}} 
P_{0}
=
P_{0}
\sum_{g}
|U_{i_r;k_{\rm F}g}|^{2}
\ff s_{\rm F}(i_{r}) 
\: , 
\ee
where 
\be
\ff s_{\rm F}(i_{r}) 
=
\frac12 \sum_{\sigma\sigma'} 
c^\dagger_{k_F\alpha\sigma}(i_r)
\ff \sigma_{\sigma\sigma'}
c_{k_F\alpha\sigma'}(i_r) \Big|_{\alpha=1}
\label{eq:effspin}
\ee
is the conduction-electron spin on the (spin-degenerate) Fermi orbital $| {\rm F}, i_{r} , \sigma \rangle \equiv c^\dagger_{k_F,\alpha=1,\sigma}(i_r) |{\rm vac.} \rangle$.
Explicitly, we have
\be
| {\rm F}, i_{r} , \sigma \rangle
=
\frac{1}{
\sqrt{\sum_{g} |U_{i_r;k_{\rm F}g}|^{2}}
}
\sum_{g}
U^{\dagger}_{k_{\rm F}g; i_{r}} 
\sum_{i} U_{i; k_{F}g} | i , \sigma \rangle
\: .
\label{eq:efforbital}
\ee

The goal of the transformation (\ref{eq:str}) is to cast the effective Hamiltonian in a spin-only form.
With the definition of the effective coupling, 
\be
J_{\rm eff}(i_{r}) = J \, \sum_{g} | U_{i_r;k_{\rm F} g} |^{2} \;,
\label{eq:effj}
\ee
Eq.~(\ref{eq:effmodel}) follows as
\be
H_{\rm eff} = P_{0} \sum_{r=1}^{R} 
J_{\rm eff}(i_{r})  
\ff s_{\rm F}(i_{r}) \cdot 
\ff S_{r}
\: .
\label{eq:effh}
\ee
Formally, the effective Hamiltonian has the structure of a central-spin model with the subspace of conduction-electron states at the Fermi level as the ``central'' degrees of freedom. 
However, the low-energy physics at weak $J$ decisively depends on different orbitals $| {\rm F}, i_{r} , \sigma \rangle$ that define the conduction-electron spins $\ff s_{\rm F}(i_{r})$ to which the impurity spins $\ff S_{r}$ are coupled.
These orbitals are degenerate one-particle eigenstates of $H_{0}$ with energy $\varepsilon_{\rm F}$ which are delocalized over the entire lattice.
Since $| {\rm F}, i_{r} , \sigma \rangle$ is in general different for each impurity spin $\ff S_{r}$, Eq.\ (\ref{eq:effh}) may represent an unconventional central-spin model.

%%%%%%%%%%%%%%%%%%%%%%%%%%%%%%%%%%%%
\begin{figure}[b]
\centerline{\includegraphics[width=0.7\columnwidth]{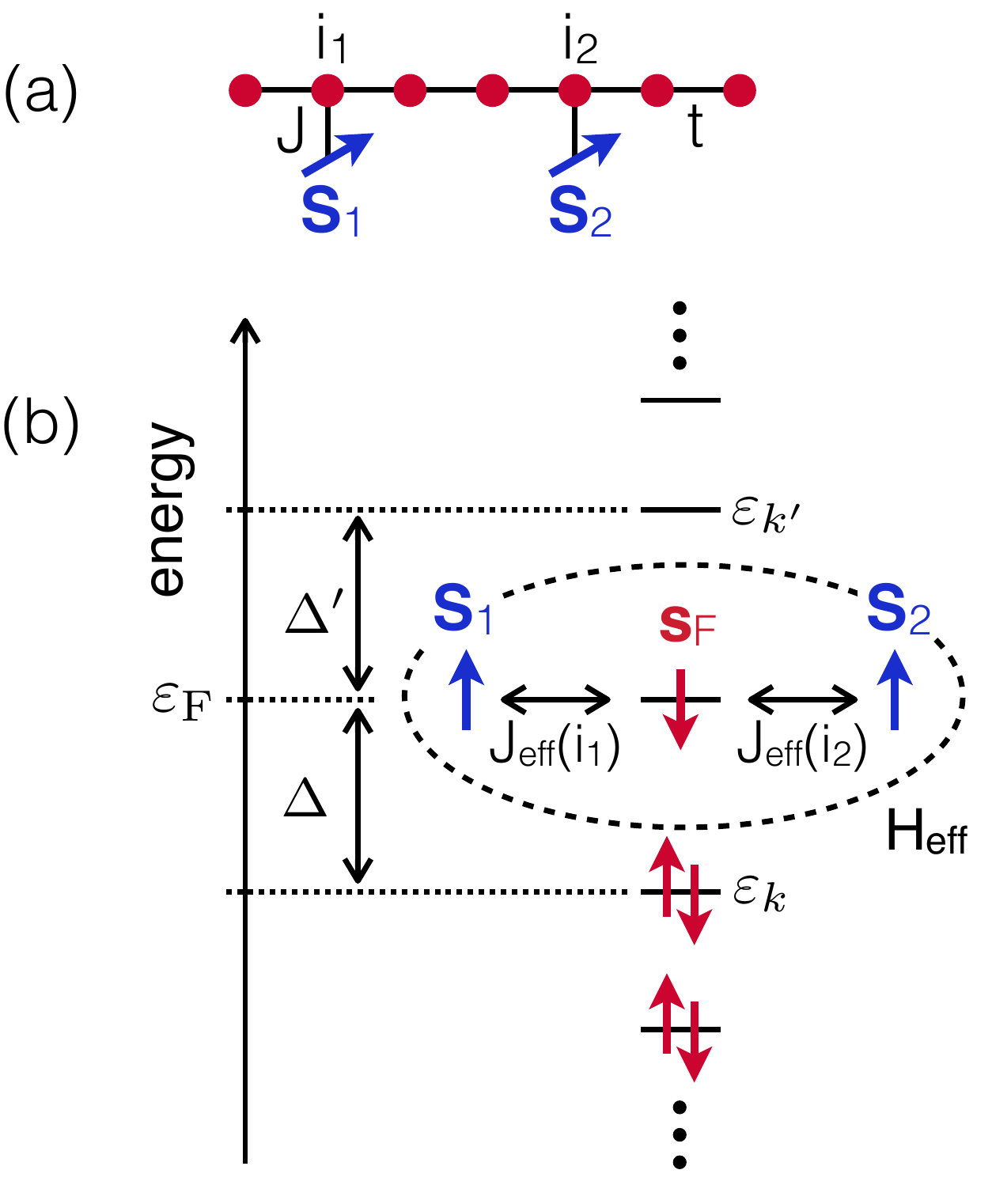}}
\caption{
(Color online) Schematic illustration of the effective low-energy theory:
(a) Multi-impurity Kondo model with hopping $t$ and antiferromagnetic exchange coupling $J$. 
Example with two impurity spins $\ff S_{1}$ and $\ff S_{2}$ coupled to the conduction-electron spins at sites $i_{1}$ and $i_{2}$ of a one-dimensional chain.
(b) One-particle eigenenergies $\varepsilon_{k}$ for $J=0$.
A single electron occupies the non-degenerate one-particle state at the Fermi energy $\varepsilon_{\rm F}$. 
$\Delta$ ($\Delta'$) is the finite-size gap.
For $0<J\ll t$ the effective low-energy model $H_{\rm eff}$ (dashed line) is given by a central-spin model with effective couplings $J_{\rm eff}(i_{r})$ between $\ff S_{r}$ and the spin $\ff s_{\rm F}$ of the delocalized one-particle state at $\varepsilon_{\rm F}$.
Only a single screening channel is available in this case.
}
\label{fig:heff}
\end{figure}
%%%%%%%%%%%%%%%%%%%%%%%%%%%%%%%%%%%%

Depending on the details of the geometry and the electronic structure, the values of $J_{\rm eff}(i_{r})$ and also the definitions of $\ff s_{\rm F}(i_{r})$ in Eq.~(\ref{eq:effh}) can be very different. 
Several distinct situations can arise, which we discuss now in turn.

Firstly, the coupling constant $J_{\rm eff}(i_{r})$ for a particular impurity spin $\ff S_{r}$ can vanish if $|U_{i_r;k_{\rm F} g} |^{2} = 0$ for all $g$ (i.e., each conduction-electron wave function $U_{i;k_{\rm F} g}$ has a node at the site $i_{r}$). In this case, there is no finite-size Kondo screening of the impurity at site $i_r$ on the linear-in-$J$ scale, and second-order perturbation theory including RKKY couplings must be employed to understand the lowest-energy physics. By contrast, an impurity at another site $i_{r'}$ with finite $J_{\rm eff}(i_{r'})$ can still be screened by forming a delocalized spin-singlet state with a conduction electron at $\varepsilon_{\rm F}$.
This situation was discussed in Ref.\ \onlinecite{SGP12} for the case with $G(k_{\rm F})=1$ but $\Gamma=2$ (a single level at the Fermi energy, with an odd total number of electrons), in the context of a one-dimensional Kondo model at half-filling with up to three impurities. 

Secondly, we consider the case $G(k_{\rm F})=1$ and odd $N$, as above, but for a Fermi wave function $U_{i_r;k_{\rm F}} \equiv U_{i_r;k_{\rm F} g=1}$ which is non-vanishing at all impurity sites $i_r$. 
An illustration is given by Fig.\ \ref{fig:heff} (a) and (b).
Depending on details of the geometry, the resulting coupling $J_{\rm eff}(i_{r})$ is possibly site-dependent.
Since $G(k_{\rm F})=1$, the matrices $\ff V(i_{r})$ reduce to $1\times 1$ ``matrices'' trivially ($\ff V(i_{r}) = \ff 1 = 1$), and all impurity spins $\ff S_{r}$ couple to the spin of the same orbital $| {\rm F} , \sigma \rangle \equiv | {\rm F}, i_{r} , \sigma \rangle$. 
In this case, the effective model reduces to a conventional central-spin model, 
\be
H_{\rm eff} = P_{0} 
\ff s_{\rm F}\cdot 
\sum_{r=1}^{R} 
J_{\rm eff}(i_{r})
\ff S_{r}
\: , 
\label{eq:csm}
\ee
with a single central spin $\ff s_{\rm F}$ which is independent of $i_{r}$.

This situation can be viewed as an antiferromagnetic Heisenberg model on a bipartite ``lattice'', with the central spin $\ff s_{\rm F}$ on an ``A sublattice'' site and the impurity spins $\ff S_{r}$ on the ``B sublattice'' sites. Hence, the Lieb-Mattis theorem \cite{LM62a} applies: 
The ground state has total spin $S_{\rm tot} = (R-1)/2$, i.e., there is exactly one screening channel.
For $R \ge 2$, this results in ``underscreening'' of the impurity degrees of freedom (whereby the overall ground state remains degenerate). 
The intuitive picture is that all impurity spins are antiferromagnetically aligned to the central spin, $\langle \ff s_{\rm F} \cdot \ff S_{r} \rangle < 0$, which implies ferromagnetic impurity spin correlations $\langle \ff S_{r}\cdot \ff S_{r'} \rangle > 0$.

This underscreening phenomenon can be seen most clearly in the special case of a translationally-invariant $D$-dimensional \emph{finite} lattice comprising  $L<\infty$ sites (for example, a ring geometry in $D=1$). The periodic boundary conditions in this case imply that the modulus square of the Fermi wave function $U_{i_r;k_{\rm F} g}$ is independent of the site index $i_{r}$. 
Hence 
\be
H_{\rm eff} = P_{0} 
J_{\rm eff}
\ff s_{\rm F} \cdot
\sum_{r=1}^{R} 
\ff S_{r}
\: . 
\label{eq:csm1}
\ee
The model can then be straightforwardly solved by using the standard rules for adding angular momenta:
The ground state of $H_{\rm eff}$ in Eq.\ (\ref{eq:csm1}) with ground-state energy $E_{\rm eff, 0} = - J_{\rm eff} (R+1) / 4$ is obtained for maximum total impurity-spin quantum number $S_{\rm imp} = \nicefrac R2$ and for total spin quantum number $S_{\rm tot} =S_{\rm imp} - \nicefrac 12$, consistent with the Lieb-Mattis theorem, and is $(2S_{\rm tot}+1)$-fold degenerate. 
The $z$-components of $\ff s_{\rm F}$ and of the total impurity spin $\ff S_{\rm imp} = \sum_{r} \ff S_{r}$ in the ground state with maximal magnetic quantum number $M_{\rm tot} = S_{\rm tot}$ are given by $\langle s_{{\rm F,}z} \rangle = - \nicefrac12 + \nicefrac {1}{R+1}$ and $\langle S_{{\rm imp}, z}  \rangle = \nicefrac R2 - \nicefrac{1}{R+1}$, respectively.

Finally, and most interesting, we discuss the case where each impurity spin $\ff S_{r}$ couples to a \emph{different} conduction-electron spin, corresponding to the Fermi orbital $| {\rm F}, i_{r} , \sigma \rangle$.
The inner product of the Fermi orbitals is given by
\be
\langle {\rm F}, i_{r'} , \sigma' | {\rm F}, i_{r} , \sigma \rangle 
=
\frac{
\delta_{\sigma\sigma'}
\sum_{g}
U_{i_{r'};k_{\rm F}g}
U^{\dagger}_{k_{\rm F}g; i_{r}}
}{ 
\sqrt{\sum_{g} |U_{i_{r'};k_{\rm F}g}|^{2}} \sqrt{\sum_{g} |U_{i_r;k_{\rm F}g}|^{2}} 
}
\: .
\label{eq:inner}
\ee
Suppose that all orbitals $| {\rm F}, i_{r} , \sigma \rangle$ are mutually orthogonal and 
that the number of impurity spins $R$ is equal to the degeneracy of the Fermi energy $G(k_{\rm F})$.
In this case, the effective model (\ref{eq:effh}) describes a Heisenberg model which decomposes into a set of $R=G(k_{\rm F})$ decoupled antiferromagnetic subsystems, each consisting of an impurity spin $\ff S_{r}$ and the corresponding conduction-electron spin $\ff s(i_{r})$.
The ground state is then a total spin singlet --- there are as many screening channels as impurity spins, and one has ``perfect'' screening. 
The degeneracy of the one-particle eigenenergies at the Fermi level, and mutual orthogonality of the states $| {\rm F}, i_{r} , \sigma \rangle$, typically results from spatial symmetries of the system. 
A concrete example of this is discussed in Sec.\ \ref{sec:dmrg}.

The case $R>G(k_{\rm F})$ corresponds to underscreening, since the number of available orthogonal Fermi orbitals is insufficient to form independent singlets with all impurity spins. 
Consequently, some of the impurity spins do not couple to the conduction-electron systems (to first order in $J$), leaving a spin-degenerate ground state.

On the other hand, if $R<G(k_{\rm F})$, there are more screening channels than necessary.  
However, each impurity spin $\ff S_{r}$ is coupled to at most one conduction-electron orbital $| {\rm F}, i_{r} , \sigma \rangle$, precluding the frustrated ``overscreening'' of any given impurity. In this case, some of the Fermi electrons simply remain decoupled from the impurities, giving rise to a trivial ground-state degeneracy.
  
If the inner product (\ref{eq:inner}) is finite, the situation is more complicated because the creators and annihilators defining the conduction-electron spins in Eq.\ (\ref{eq:effh}) do not refer to orthogonal orbitals. Generally this leads to an unconventional central-``spin'' model,  where the Fermi electrons can interact with more than one impurity.
This case requires a different numerical approach, and will be discussed in a separate paper. \cite{HP15}

%--------------------------------------------------------------------------------------------------------------
\section{Crossovers between strong coupling and Kondo or RKKY regimes}
\label{sec:dmrg}
%--------------------------------------------------------------------------------------------------------------

Having explored the expected physics of finite-sized magnetic nanostructures on general grounds, we turn now to a specific example with $R=2$ impurity spins coupled to a conduction-electron system with doubly-degenerate orbitals at the Fermi energy ($G(k_{\rm F})=2$). 

Going beyond the analysis of the previous section, here we also consider the various crossovers expected to arise upon increasing the strength of the exchange coupling, $J$. These crossovers are manifest in the evolution of zero-temperature static spin-spin correlation functions, which we calculate numerically. In particular, when the two impurities are locked together into a spin-singlet, we expect $\langle \ff S_{1} \cdot \ff S_{2} \rangle = - \nicefrac{3}{4}$ for the inter-impurity spin correlation. 
This situation pertains when the impurities are coupled by an effective non-local antiferromagnetic RKKY interaction. On the other hand, if the Kondo effect dominates (and provided both impurity spins can be screened), we expect $\langle \ff S_{r} \cdot \ff s_{\rm tot} \rangle = - \nicefrac{3}{4}$ for the correlation between each impurity spin $r=1,2$ and the total conduction-electron spin $\ff s_{\rm tot} = \sum_{i} \ff s_{i}$. The inter-impurity correlation should vanish in this case, $\langle \ff S_{1} \cdot\ff S_{2} \rangle = 0$, signaling that the impurities are decoupled from each other. 
%Indeed, when the overall ground state is a singlet (whether it be Kondo or RKKY), spin conservation implies directly that $\langle \ff S_{1} \cdot \ff S_{2} \rangle + \langle \ff S_{r} \cdot \ff s_{\rm tot} \rangle = -\tfrac{3}{4}$.

The question of which process dominates is a subtle one, since the RKKY coupling cannot be treated as an independent parameter but rather depends on the Kondo coupling $J$. 
Furthermore, besides the RKKY coupling $J_{\rm RKKY}$ and the Kondo temperature $T_{\rm K}$, the finite-size gap $\Delta$ represents a third energy scale. The competition between these is the focus of our present study.

%%%%%%%%%%%%%%%%%%%%%%%%%%%%%%%%%%%%
\begin{figure}[t]
\centerline{\includegraphics[width=0.98\columnwidth,angle=0]{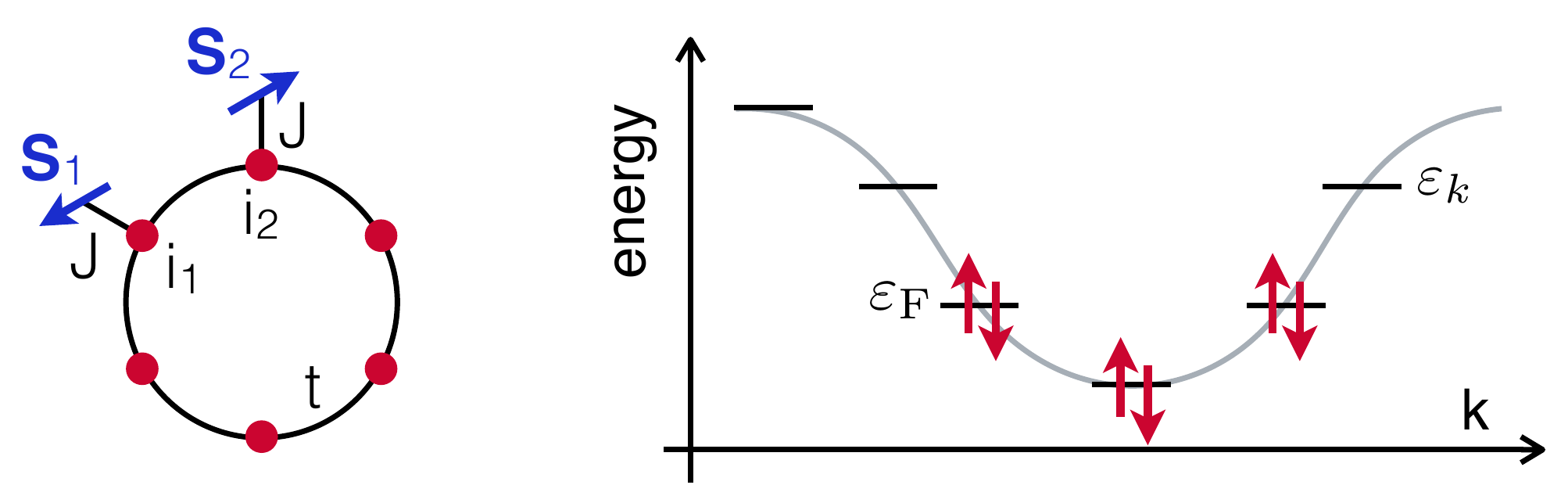}}
\caption{
(Color online) 
{\em Left:} Two-impurity Kondo model on a finite one-dimensional lattice with periodic boundary conditions. 
Off-resonance case with $L=4m+2$ (illustrated for $m=1$).
{\em Right:} Schematic picture of the one-particle eigenenergies $\varepsilon_{k}$ for $J=0$.
Eigenenergies (except for the lowest and highest) are two-fold degenerate, $G(k_{\rm F})=2$. 
$\varepsilon_{\rm F}$ denotes the highest occupied one-particle energy. 
At half-filling $N=L$ the one-particle states at $\varepsilon_{\rm F}$ are fully occupied.
}
\label{fig:sysoff}
\end{figure}
%%%%%%%%%%%%%%%%%%%%%%%%%%%%%%%%%%%%

As a concrete example, we now consider a one-dimensional lattice with a finite, but possibly large number of sites $L$, to which two spin-$\tfrac{1}{2}$ impurities are coupled. This two-impurity Kondo model can be solved for arbitrary coupling strength $J$ by numerical means, as discussed further below.

Specifically, we assume a non-zero hopping $t_{ij} = - t$ between nearest neighbors $i$ and $j$ of a one-dimensional lattice, and set $t=1$ to fix the energy scale. Two impurity spins $\ff S_{1}$ and $\ff S_{2}$ are coupled to \emph{nearest neighbor} sites $i_{1}$ and $i_{2}$.
We realize two-fold degenerate Fermi orbitals, $G(k_{\rm F})=2$, by imposing periodic boundary conditions (i.e.\ a ring geometry). The model is studied at zero temperature and half-filling ($N=L$ electrons). 

The hopping matrix is diagonalized [c.f.\ Eq.\ (\ref{eq:umatrix})] by discrete Fourier transformation, 
\be
  U_{i; kg} = \frac{1}{\sqrt{L}} e^{\pm i k R_{i}} \; ,
\label{eq:fourier}
\ee
with the ``$+$'' sign for $g=1$ and ``$-$'' sign for $g=2$, wave vector $k = - (\pi / a) + n \Delta k$ (where $n=0,1,...,L-1$), and position vector $R_{i} = i \cdot a$ for site $i=0,1,...,L-1$. 
Here, $a$ is the lattice constant and $\Delta k = 2\pi / (aL)$. 
Note that the effective couplings (\ref{eq:effj}) do not vanish since $|U_{i; kg}|^{2} \ne 0$. 
The two-fold degenerate one-particle energies are given by 
\be
   \varepsilon_{k} = - 2 t \cos( ka ) 
   \: .
\label{eq:dispersion}
\ee
This results in a finite-size gap at the Fermi energy of $\Delta = 2t \sin(2\pi/L) \propto 1/L$ for large $L$. 

For moderate system sizes up to $L=100$ and intermediate coupling strengths, numerical results for the static $T=0$ spin-spin correlators $\langle \ff S_{1} \cdot\ff S_{2} \rangle$ and $\langle \ff S_{r} \cdot \ff s_{\rm tot} \rangle$ can be accurately obtained using the density-matrix renormalization group (DMRG).\cite{Whi92} We implement here a standard scheme based on matrix-product states and matrix-product operators; see Ref.\ \onlinecite{TSRP12} for a brief discussion of the algorithm, and Ref.\ \onlinecite{Sch11} for a general overview of matrix-product-state techniques. 

Usually, DMRG is formulated for and applied to one-dimensional systems with open boundaries.
To employ the standard algorithm for the present case with periodic boundaries, we first consider the open chain consisting of $L$ sites and connected by nearest-neighbor hopping terms $t_{i, i+1}$ and $t_{i+1, i}$ for $i=1,...,L-1$ as usual.
The chain is folded in half and connected at the open end with the missing hopping terms $t_{1,L}$ and $t_{L,1}$ to generate a new half-length chain with new ``sites'' consisting of pairs $(i,L-i+1)$ of original sites.
We let the standard DMRG algorithm operate on the new ``sites''.
The approach avoids long-range hopping terms in the Hamiltonian (which are unfavorable for the scaling of DMRG), at the cost of an enlarged local Hilbert space. We address larger system sizes and weaker $J$ in Sec.~\ref{sec:nrg} using an alternative approach based on the numerical renormalization group.\cite{Wil75,nrg_rev}

%%%%%%%%%%%%%%%%%%%%%%%%%%%%%%%%%%%%
\begin{figure}[t]
\centerline{\includegraphics[width=0.95\columnwidth]{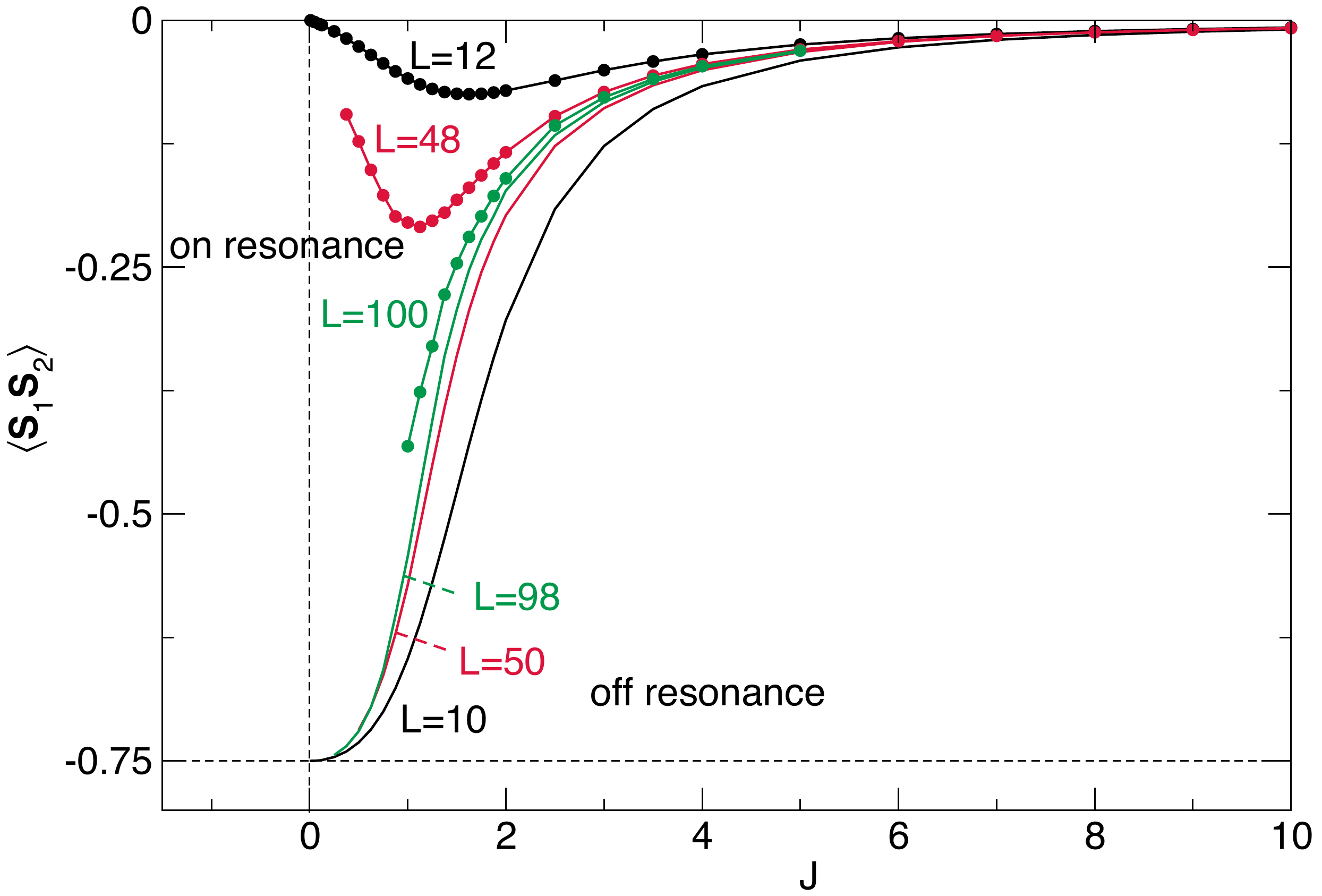}  }
\centerline{\includegraphics[width=0.95\columnwidth]{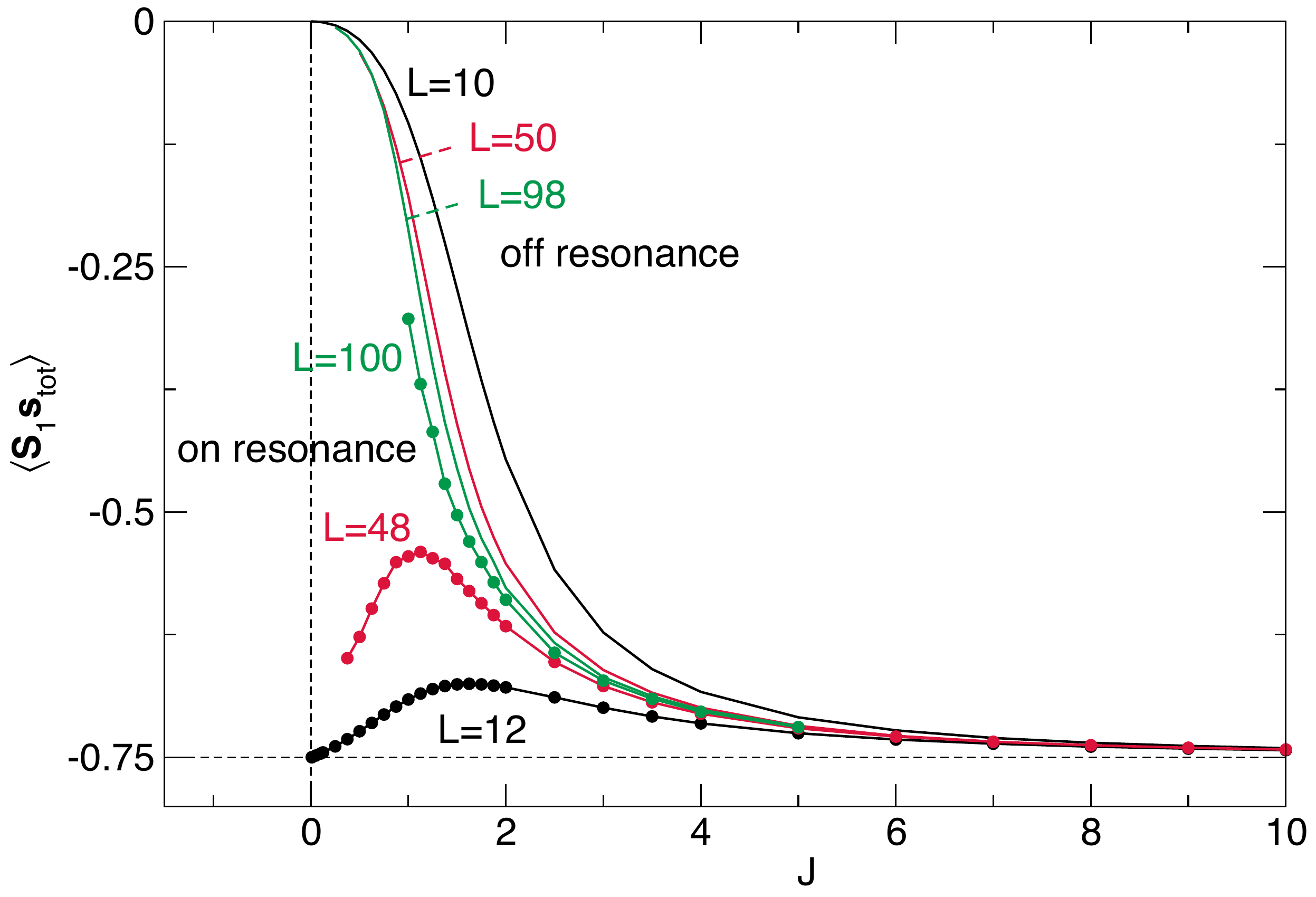}  }
\caption{
(Color online)
Evolution of the spin-spin correlation functions $\langle \ff S_{1} \cdot\ff S_{2} \rangle$ (top) and $\langle \ff S_{1}\cdot \ff s_{\rm tot} \rangle$ (bottom) with $J$, as obtained from DMRG for various system sizes $L$. 
Off-resonance cases for $L=4m+2$ (lines), on-resonance cases for $L=4m$ with integer $m$ (dots). 
See Fig.\ \ref{fig:sysoff} for a schematic of the setup. 
}
\label{fig:dmrg}
\end{figure}
%%%%%%%%%%%%%%%%%%%%%%%%%%%%%%%%%%%%

We discuss first off-resonance situations, realized at half-filling for $L=4m+2$ with integer $m$ --- see Fig.\ \ref{fig:sysoff} for an illustration. 
Here the ground state for $J=0$ is the non-degenerate Fermi sea, and all one-particle states at $\varepsilon_{\rm F}$ are completely filled. 
Recall from Sec.~\ref{sec:theory} that no linear-in-$J$ finite-size Kondo effect can arise in the off-resonance case, because all Fermi orbitals are completely occupied. Instead, the effective low-energy physics in the limit $J\to 0$ is described by the RKKY model, Eq.\ (\ref{eq:rkky}).
The sign of the RKKY exchange coupling depends on the distance $d \equiv | i_1 - i_2 |$ between the impurity spins; for the present case of a half-filled one-dimensional system, $J_{\rm RKKY}\propto (-1)^{d+1}$. In particular, for nearest neighbors ($d=1$) the coupling is \emph{antiferromagnetic}, $J_{\rm RKKY} > 0$. 

Fig.\ \ref{fig:dmrg} shows DMRG results for the inter-impurity correlations $\langle \ff S_{1}\cdot \ff S_{2} \rangle$ and for the correlation between one of the impurity spins and the total conduction-electron spin $\langle \ff S_{1}\cdot \ff s_{\rm tot} \rangle$. Note that reflection symmetry implies $\langle \ff S_{1}\cdot \ff s_{\rm tot} \rangle = \langle \ff S_{2}\cdot \ff s_{\rm tot} \rangle$. 
Results have been obtained for different system sizes $L=10,50,98$. 
Fig.\ \ref{fig:dmrg} demonstrates that the impurities are indeed effectively decoupled from the conduction-electron system for $J  \to 0$, as is indicated by the vanishing correlation function $\langle \ff S_1\cdot \ff s_{\rm tot} \rangle \to 0$. 
The antiferromagnetic RKKY exchange, coupling impurity spins $\mathbf S_1$ and $\mathbf S_2$, is dominant at weak coupling. This results in the formation of a non-local RKKY singlet, with $\langle \ff S_1 \cdot\ff S_2\rangle \to - \nicefrac34$ as $J \to 0$. 

With increasing $J$ one observes a smooth crossover from this RKKY regime to a strong-coupling regime where the two impurity spins are separately screened by the conduction electrons. 
This is reflected by a vanishing inter-impurity correlation $\langle \ff S_1 \cdot\ff S_2\rangle \to 0$ as $J \to \infty$, and by Kondo correlations $\langle \ff S_r\cdot \ff s_{\rm tot} \rangle \to -\nicefrac34$. 
For large $J$, the Kondo effect simply reduces to the formation of entirely local singlets, i.e., we find $\langle \ff S_r \cdot\ff s_{i_{r}} \rangle \to - \nicefrac34$ for the \emph{local} Kondo correlation function (not shown in the figure).

The system-size dependence of the crossover from RKKY to Kondo physics is quite regular in the off-resonance case, and the results for $L=98$ are rather representative of the physics in the thermodynamic limit $L\to \infty$. This is demonstrated explicitly and discussed further in Sec.~\ref{sec:nrg}. 
The off-resonance behavior here can be understood from the conventional Doniach scenario;\cite{Don77} it is associated with a characteristic coupling strength $J_{\rm D}$ separating RKKY and Kondo regimes.

\begin{figure}[t]
\centerline{\includegraphics[width=0.98\columnwidth,angle=0]{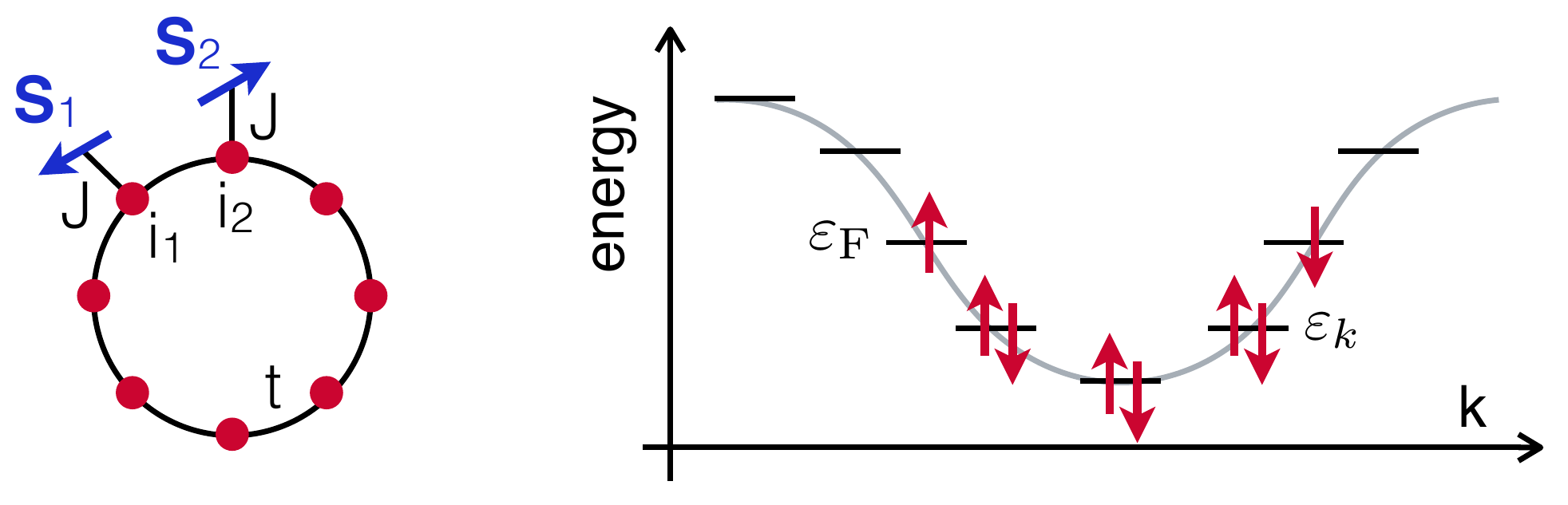}}
\caption{
(Color online) 
Same as Fig.\ \ref{fig:sysoff}, but for an on-resonance case with $L=4m$ (illustrated here for $m=2$).
At half-filling $N=L$, the one-particle states at $\varepsilon_{\rm F}$ are partially occupied.
}
\label{fig:syson}
\end{figure}

However, the on-resonance situation (realized here for $L=4m$, see Fig.\ \ref{fig:syson}) is more complex, since there one expects a second crossover on the scale of $J_{\Delta}$ to a regime dominated by the linear-in-$J$ finite-size Kondo effect. We explore this physics now in the context of full DMRG results with $L=12,48, 100$ presented in Fig.\ \ref{fig:dmrg}. 

First, we comment briefly on the strong-coupling limit $J\gg J_{\rm D}, J_{\Delta}$. Here the physics is essentially the same as that arising in the off-resonance case, both being dominated by strong \emph{local} spin correlations $\langle \ff S_{r}\cdot \ff s_{i_{r}} \rangle$. Details of the electronic structure close to the Fermi level then become irrelevant. This can be seen directly by comparing the off- and on-resonance situations (lines vs.\ dots) in Fig.\ \ref{fig:dmrg}. For example, the same basic behavior is found for all $J>1$ with $L=98$ and $L=100$ (green lines and dots); and agreement becomes essentially quantitative for $J>3$ or so.

%%%%%%%%%%%%%%%%%%%%%%%%%%%%%%%%%%%%
%%%%%%%%%%%%%%%%%%%%%%%%%%%%%%%%%%%%

%%%%%%%%%%%%%%%%%%%%%%%%%%%%%%%%%%%%
\begin{figure}[b]
\centerline{\includegraphics[width=0.55\columnwidth,angle=0]{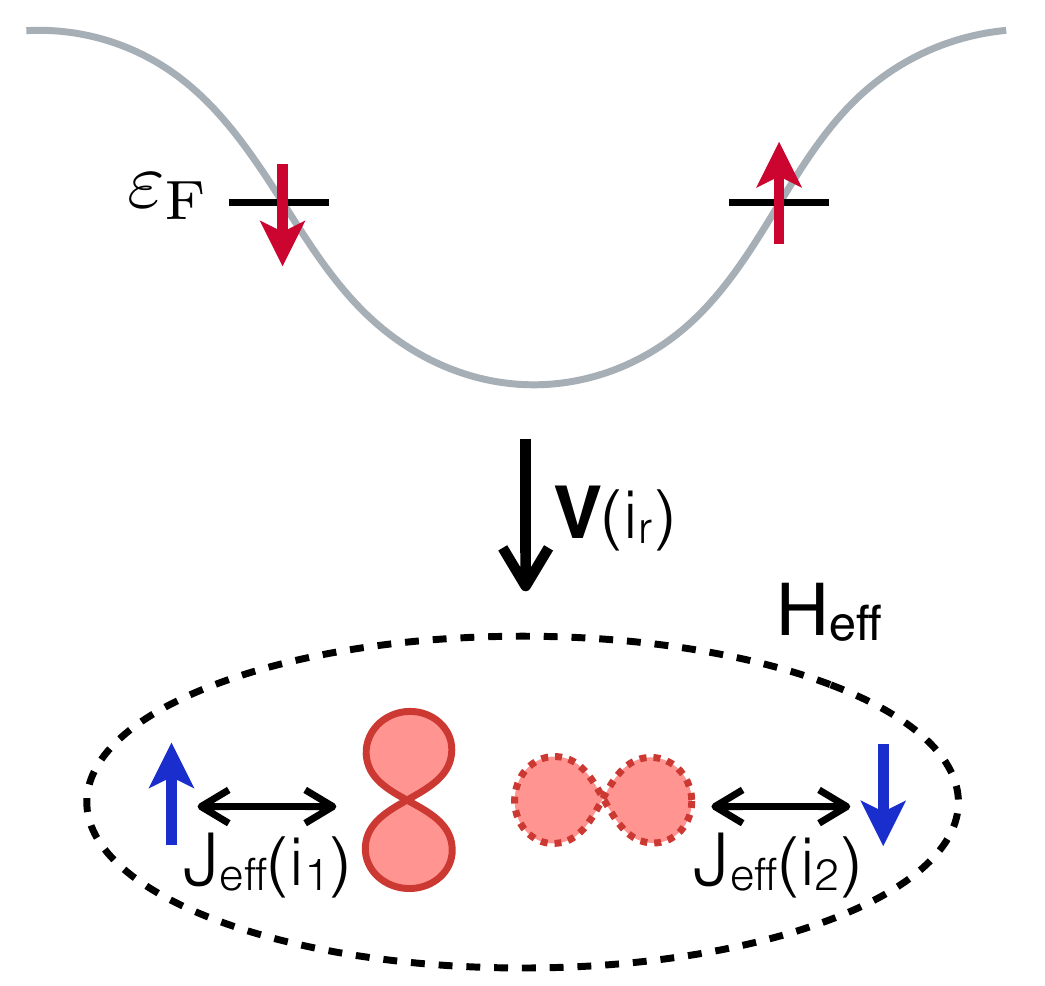}}
\caption{
(Color online) 
Schematic picture of the effective low-energy model of the system depicted in Fig.\ \ref{fig:syson}.
The site-dependent unitary transformation $\ff V(i_{r})$ generates orthogonal Fermi orbitals. 
}
\label{fig:syseff}
\end{figure}
%%%%%%%%%%%%%%%%%%%%%%%%%%%%%%%%%%%%

By contrast, on- and off-resonance cases are qualitatively very different at weak coupling. Here, the low energy details of the conduction-electron system become important. 
In the on-resonance case, the one-particle states at the Fermi energy are half-filled, as sketched in Fig.\ \ref{fig:syson}. The resulting degeneracy of the many-body ground state at $J=0$ gives rise to a linear-in-$J$ Kondo scale, which dominates as $J \to 0$. The effective low-energy model is then given by Eq.\ (\ref{eq:effh}).

Indeed, our DMRG results are indicative of a vanishing inter-impurity correlation $\langle \ff S_{1}\cdot \ff S_{2} \rangle \to 0$ as $J\to 0$, implying that the RKKY interaction is inoperative at weak coupling. 
Strong Kondo correlations do however appear to develop, with $\langle \ff S_{r}\cdot \ff s_{\rm tot} \rangle \to -\nicefrac34$ as $J\to 0$. This is seen clearly from the numerics with $L=12$ (and results for $L=48$ are highly suggestive).  However, we were unable to stabilize well-converged DMRG results for larger systems and smaller $J$, due to the need to resolve small energy scales $\Delta \sim 1/L$, and the development of highly extended entangled ground states. Nevertheless, the results are consistent with our expectation that the ground state for $L=4m$ is an overall spin-singlet, with each impurity being separately screened. 

In contrast to the strong-coupling (large $J$) limit, where the screening can be visualized as two spatially separated and almost perfectly localized Kondo singlets, for small $J<J_{\Delta}$ the ``Kondo clouds'' extend over the \emph{entire} system (i.e.\ they have extent $L$). This implies the existence of two different screening channels. Similarly, in the thermodynamic limit $L\rightarrow \infty$, it was pointed out in Ref.~\onlinecite{mitchell2015multiple} that perfect and essentially \emph{independent} Kondo screening can take place for nearby impurities, even though their Kondo clouds are extended and substantially overlapping in real-space. The difference, however, is that as $L\rightarrow \infty$ the conduction-electron system becomes gapless $\Delta\rightarrow 0$, and the Kondo cloud for a single impurity then has extent 
$\xi_{\rm K}\sim v_{\rm F}/ T_{\rm K}$ 
(with $v_{\rm F}$ the Fermi velocity), which can be understood from a renormalization group perspective.\cite{MBB11}
 
The DMRG results are fully consistent with the implications of the effective low-energy Hamiltonian (\ref{eq:effh}). 
For $J<J_\Delta$ the two impurities interact with a central region given by the one-particle states at $\varepsilon_{\rm F}$ and the two Fermi electrons. 
To determine whether the impurities couple to orthogonal states or not, we evaluate Eq.\ (\ref{eq:inner}) using the wave functions given by Eq.\ (\ref{eq:fourier}): 
\be
\langle {\rm F}, i_{1} , \sigma' | {\rm F}, i_{2} , \sigma \rangle 
=
\delta_{\sigma\sigma'}
\cos \left[
k_{\rm F} 
(
R_{i_1} - R_{i_2}
)
\right]
\: .
\label{eq:overlap}
\ee
Note the different dependence on the distance as compared to the RKKY coupling, which is $J_{\rm RKKY} \propto \cos[2k_F(R_{i_1} - R_{i_2})]$. 
At half-filling, the Fermi wave vectors are $k_{\rm F} = \pm \pi / 2a$.
With $| R_{1} - R_{2} | = a \cdot d$, we have 
$\langle {\rm F}, i_{1} , \sigma | {\rm F}, i_{2} , \sigma \rangle = \cos ( \pi d /2)$.
In the case of nearest neighboring impurities ($d=1$) the matrix element vanishes, implying that the impurity spins couple to orthogonal channels, and are screened separately (see Fig.\ \ref{fig:syseff}).

To summarize these findings:
For very large $J$, each impurity is Kondo-screened by conduction-electron states which are highly localized in real-space. As $J$ is decreased, the Kondo screening becomes more delocalized. As $J$ is reduced further, below $J_{\rm D}$, there is a crossover to an inter-impurity RKKY singlet state. In the off-resonance case, the RKKY interaction continues to dominate as $J\to 0$. But in the on-resonance case, a second crossover arises on the scale of $J_{\Delta}$ to a state characterized by the finite-size Kondo effect, with orthogonal Kondo clouds spreading over the entire lattice. 
Provided $J_{\Delta}\ll J_{\rm D}$, these regimes will be distinct, and re-entrant Kondo physics should be seen. 

Since $\Delta \sim 1/L$ in one dimension and thus $J_{\Delta} \sim 1 / \ln L$, one generally expects that two successive crossovers will be cleanly observed upon decreasing $J$ for \emph{larger} system sizes. This is of course physically sensible, since in the $L\rightarrow \infty$ limit, the second (finite-size Kondo) crossover is pushed to $J\rightarrow 0$, and becomes unobservable. By contrast, for small $L$ (see e.g.\ DMRG results for $L=12$ and $48$ in Fig.~\ref{fig:dmrg}), $J_{\Delta}$ and  $J_{\rm D}$ are not well-separated. Then, the RKKY regime is never fully realized, and $\langle \ff S_{1} \cdot\ff S_{2} \rangle > - \nicefrac34$ for all $J$.

Finally, we touch on the physics arising when the impurity spins are separated by an even distance, $d$. 
Here, the modulus of the overlap (\ref{eq:overlap}) is unity, implying that both impurity spins couple to the spin in the same Fermi orbital.  
The ground state in this case is a tensor product of the (underscreened) doublet ground state of the corresponding central-spin model, and the Kramers doublet of the Fermi sea (with the remaining unpaired electron at $\varepsilon_{\rm F}$).

%--------------------------------------------------------------------------------------------------------------
\section{Approaching the thermodynamic limit}
\label{sec:nrg}
%--------------------------------------------------------------------------------------------------------------

%%%%%%%%%%%%%%%%%%%%%%%%%%%%%%%%%%%%
\begin{figure*}[t]
\begin{center}
\includegraphics[width=1.5\columnwidth]{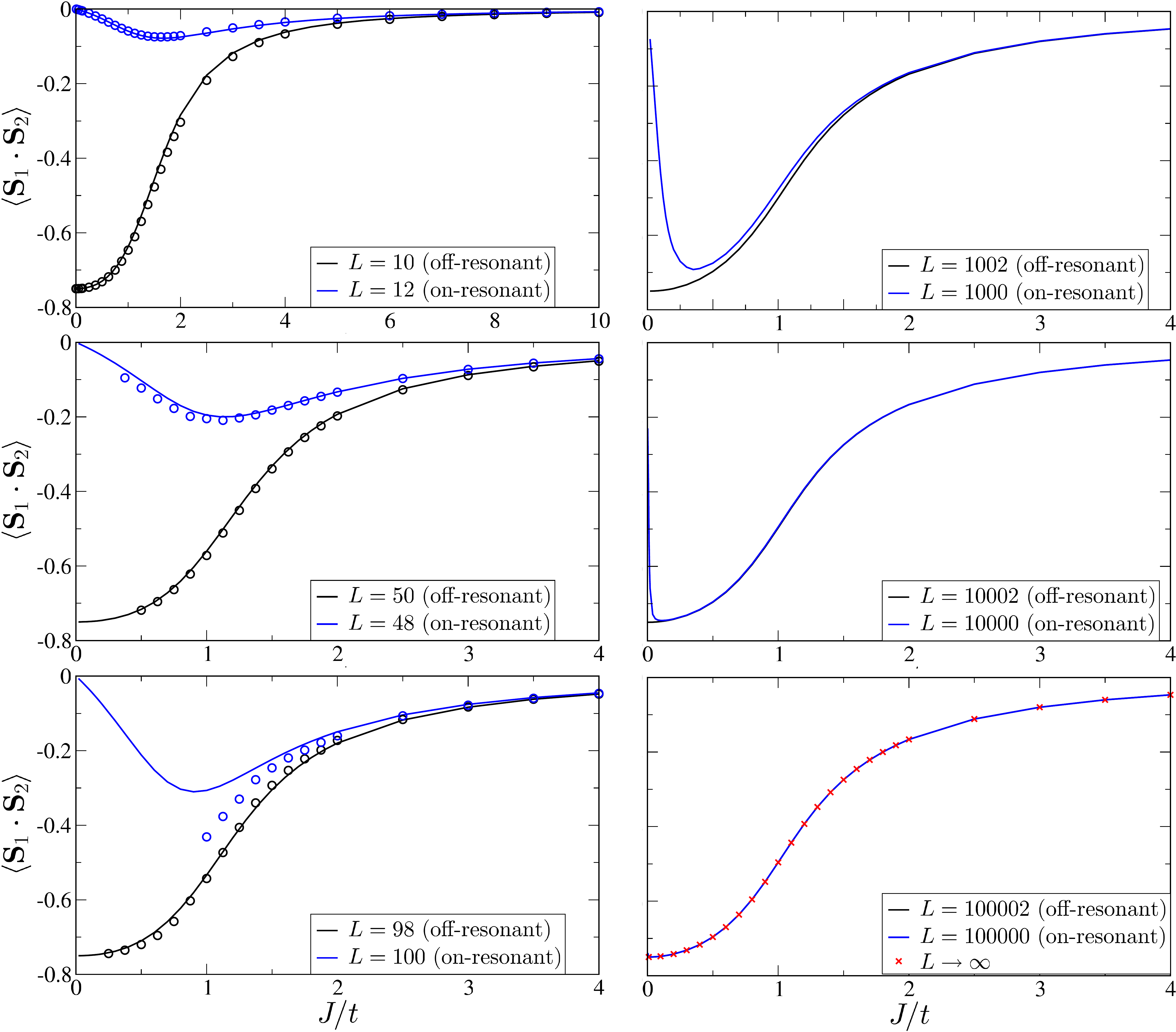}
\end{center}
\caption{
(Color online) 
Evolution of $\langle \ff S_{1} \cdot \ff S_{2} \rangle$ with $J$, as obtained from NRG for various system sizes $L$ (on-resonance cases: blue lines, off-resonance cases: black lines). 
Crosses: continuum result in the limit $L\rightarrow \infty$ obtained by standard NRG.
DMRG data from Fig.\ \ref{fig:dmrg} are shown for comparison (circles). 
}
\label{fig:s1s2}
\end{figure*}
%%%%%%%%%%%%%%%%%%%%%%%%%%%%%%%%%%%%

%%%%%%%%%%%%%%%%%%%%%%%%%%%%%%%%%%%%
\begin{figure*}[t]
\begin{center}
\includegraphics[width=0.7\textwidth]{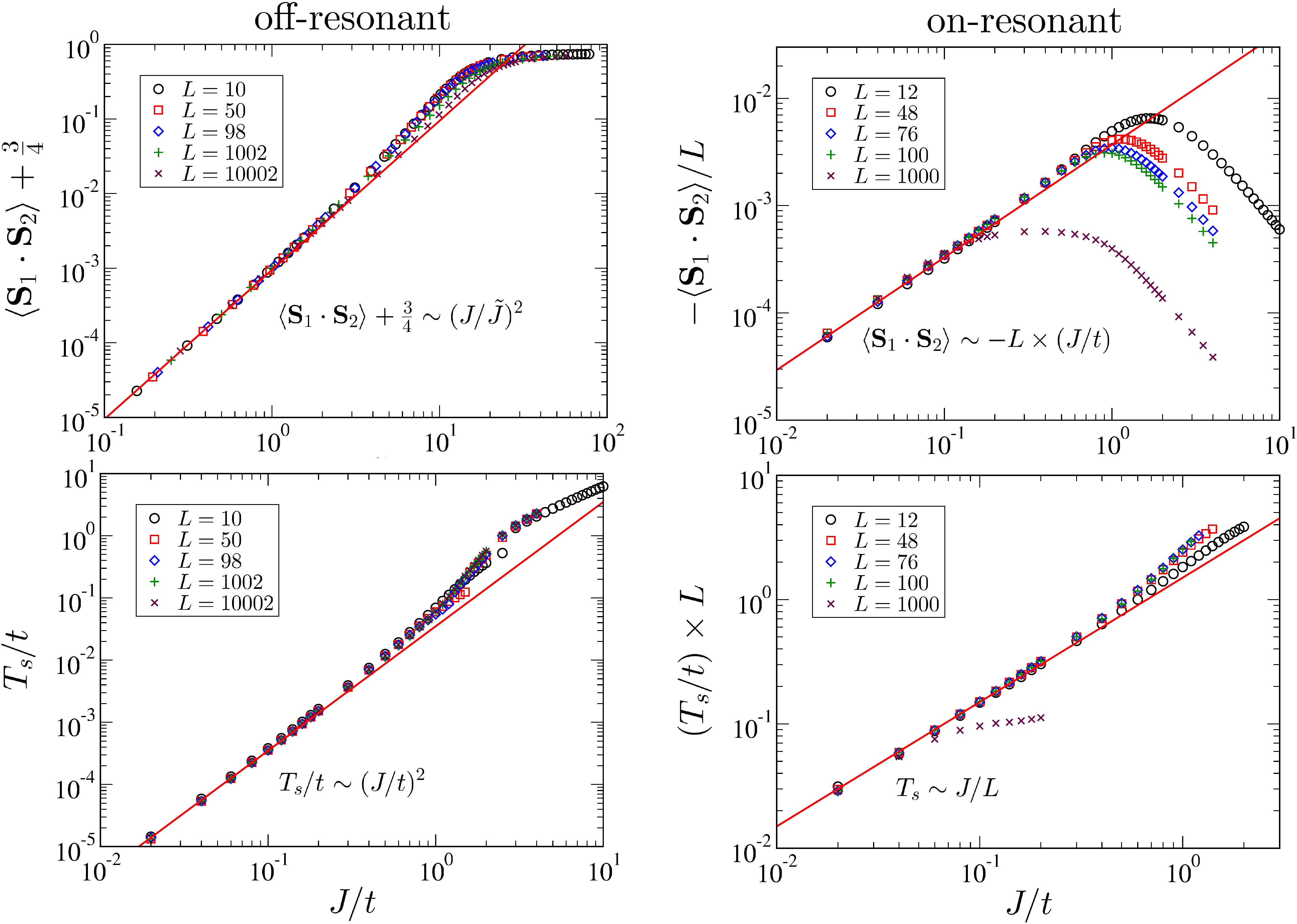}
\end{center}
\caption{
(Color online) 
{\em Upper panels:} 
$J$ dependence of $\langle \ff S_{1}\cdot \ff S_{2} \rangle$ for small $J$ for various system sizes $L$.  
{\em Lower panels:}
$J$ dependence of the low-energy scale $T_s$, corresponding to singlet formation. 
Distinct asymptotic scaling is found for off-resonance cases (left panels) and on-resonance cases (right panels). 
Results obtained by NRG.
}
\label{fig:scaling}
\end{figure*}
%%%%%%%%%%%%%%%%%%%%%%%%%%%%%%%%%%%%

For substantially larger systems, one expects that $J_{\Delta} \ll J_{\rm D}$. In this case, a well-defined RKKY regime should develop, manifest by inter-impurity spin correlations approaching $\langle \ff S_{1}\cdot \ff S_{2} \rangle \to - \nicefrac34$. 
Furthermore, the distinction between on- and off-resonance cases must become irrelevant in the thermodynamic limit $L\to \infty$ for coupling strengths $J \gg J_{\Delta} \sim 1/ \ln L \to  0$. Then, the conduction-electron system becomes gapless, and it is unimportant that states precisely at the Fermi energy are either fully or partially occupied.

As discussed in Sec.~\ref{sec:dmrg}, larger systems and weaker couplings cannot in practice be investigated with DMRG. To understand the evolution on increasing the system size to approach the thermodynamic limit, we now employ a different but complementary approach, based on the numerical renormalization group\cite{Wil75} (NRG). 

Traditionally, NRG is used for single impurities coupled to gapless conduction-electron systems in the thermodynamic limit (for a review, see Ref.~\onlinecite{nrg_rev}). In that case, NRG is known to provide accurate and efficient access to a wide range of thermodynamic\cite{Wil75} and dynamic\cite{anders2005real,weichselbaum2007sum} quantities, at essentially any temperature or energy scale. Although generalization of NRG to systems with several conduction-electron channels requires significantly greater computational resources,\cite{mitchell2014generalized,weichselbaum2012non} detailed results for two impurities separated in real-space on a three-dimensional lattice with $L\rightarrow \infty$ were recently obtained in Ref.~\onlinecite{mitchell2015multiple}. Note that NRG can be used for impurity models of any geometry and dimensionality, unlike DMRG which is usually confined to one spatial dimension.

To study \emph{finite} systems with NRG, we generalize the standard formulation. In particular, two features of the current problem are non-standard and must be treated differently. Firstly, the conduction-electron density of states for the finite system consists of a set of discrete poles located at $\varepsilon_k$, rather than a continuum. Secondly, in the on-resonance case of particular interest, there is a pole located at the Fermi energy, which controls the low-energy physics. At the heart of NRG lies a logarithmic discretization of the conduction-electron spectrum; in the present context, the spectrum is already discrete, and a pole at the Fermi energy cannot be resolved on the logarithmic scale. Our NRG implementation, which overcomes these issues, is discussed in the Appendix.

In the following we discuss our NRG results, obtained for the effective two-impurity, two-channel setup using a logarithmic discretization parameter $\Lambda=2$ and retaining $N_K=15000$ states at each step of the iterative process. We exploit total charge and spin-projection quantum numbers (but so-called z-trick averaging was not used). In all cases, we have verified that the results are fully converged with respect to $N_K$, indicating that the discretized model is solved exactly by NRG.

Fig.\ \ref{fig:s1s2} shows NRG results (lines) for the full $J$-dependence of the inter-impurity spin correlation $\langle \ff S_{1} \cdot \ff S_{2} \rangle$, obtained for different system sizes from $L=10$ up to $L = 10^{5}$. For small $L$, the previously discussed DMRG results are included for comparison (circle points). Off-resonance cases ($L=4m+2$) are shown in black; on-resonance ($L=4m$) in blue.

For very large system sizes, both on-resonance ($L = 10^{5}$) and off-resonance ($L = 10^{5}+2$) cases agree perfectly with results in the thermodynamic limit $L\rightarrow \infty$ (red cross points), where NRG should give numerically exact results.\cite{nrg_rev} In the other extreme, for very small system sizes $L=10$ and $L=50$ (off-resonance) or $L=12$ and $L=48$ (on-resonance), NRG agrees very well with the available DMRG data. This is a non-trivial result, since NRG involves a re-discretization of the conduction-electron spectrum, which itself only contains $L/2$ poles (see Appendix). However, for the on-resonance case $L=100$, we start to see a more significant deviation between the depth of the minimum in $\langle \ff S_{1} \cdot \ff S_{2} \rangle$ calculated using DMRG and NRG. This discrepancy can be traced to the discretization in NRG. Results obtained using different $\Lambda$ indicate that this discrepancy diminishes as $\Lambda\rightarrow 1$, where the bare model is recovered. 
We also found for a given $\Lambda>1$ that the DMRG results are recovered using NRG with a slightly larger $L$, i.e., the effect of discretization in NRG can be also be thought of as a slight renormalization of the system size $L$. 
The general trend exhibited as $L$ increases is, however, consistent with expectation from Secs.~\ref{sec:theory} and \ref{sec:dmrg}, as now discussed. 

We find that the results for the off-resonance case $L=4m+2$ converge uniformly and rapidly with increasing $L$. In all such cases, $\langle \ff S_{1}\cdot \ff S_{2} \rangle \to - \nicefrac34$ as $J \to 0$, indicating the dominant RKKY mechanism at weak-coupling.

In the on-resonance cases with finite but large $L=1000$ and $L=10000$, the spin correlations $\langle \ff S_{1} \cdot \ff S_{2} \rangle$ displayed in Fig.\ \ref{fig:s1s2} clearly reflect a dominant RKKY coupling mechanism in the intermediate-$J$ range. In particular, the minimum in $\langle \ff S_{1} \cdot \ff S_{2} \rangle$ approaches $- \nicefrac34$, symptomatic of inter-impurity singlet formation. In this regime, the corresponding value of $\langle \ff S_{r}\cdot \ff s_{\rm tot} \rangle = -\tfrac{3}{4}-\langle \ff S_{1} \cdot \ff S_{2} \rangle \approx 0$ indicates vanishing Kondo correlations.
The crossover from the strong-coupling Kondo to this RKKY regime is nearly independent of system size and is located at $J_{\rm D} \sim 1-2$. 

On the other hand, the crossover from the RKKY regime to the finite-size two-channel Kondo screened ground state takes place on the scale of $J_{\Delta} \sim 1 / \ln L$. 
For $J\gg J_{\Delta}$, the spin-spin correlations coincide with the continuum result ($L \to \infty$); only in the range $0<J<J_{\Delta}$ does one see marked deviations. However, this regime progressively shrinks to zero as $J_{\Delta} \sim 1/\ln L \to 0$, indicating that the weak-coupling limit $J\to 0$ and the thermodynamic limit $L\rightarrow \infty$ do not commute. 
For system sizes $L=100$ and smaller, $J_{\Delta}$ and $J_{\rm D}$ are of comparable size; then only incipient RKKY effects are observed. The clean crossover to a distinct RKKY regime is only observed for larger $L$.

The non-trivial weak-coupling physics is further examined in Fig.\ \ref{fig:scaling}. 
The lower panels show the characteristic energy scale $T_{s}$ which is necessary to break the singlet ground state for both off-resonance (left panel) and on-resonance (right panel) cases.  
This energy scale also controls the renormalization-group flow to the stable fixed point describing the ground state within the NRG scheme. 
Clearly, one would expect that $T_{s} \propto J^{2}$ in the off-resonance case corresponding to the effective RKKY model (\ref{eq:rkky}) and 
$T_{s} \propto J$ in the on-resonance case corresponding to the central-spin model (\ref{eq:effh}).
To check this, we have calculated with NRG the full (zero temperature) dynamical correlation function $\langle \langle \ff S_{1} ; \ff S_{2} \rangle\rangle_{\omega}$, i.e., the Fourier transform of $i\theta(t)\langle\ff S_{1}(t)\cdot \ff S_{2}(0) \rangle$, following Ref.~\onlinecite{weichselbaum2007sum}.  
As $\omega\rightarrow 0$, we find $\langle\langle \ff S_{1} ; \ff S_{2} \rangle\rangle_{\omega} \rightarrow 0$ -- the vanishing of the correlator indicates spin-singlet formation. 
At finite $\omega \sim T_s$, however, $\langle\langle \ff S_{1} ; \ff S_{2} \rangle\rangle_{\omega}$ shows a peak resulting from spin-flip fluctuations. 
Hence, $T_s$ can be interpreted as the energy scale for singlet formation. 
From the lower panels of Fig.\ \ref{fig:scaling} we can read off the following asymptotic behavior:
\begin{eqnarray}
&&
T_s \overset{J\rightarrow 0}{\sim} J^{2} ~\text{(off resonance)} 
\; ,
\nonumber \\
&&
T_s \overset{J\rightarrow 0}{\sim} J ~\text{(on resonance)} 
\; ,
\label{eq:Ts_asymptotes}
\end{eqnarray}
which nicely agrees with our expectations, and confirms the perturbative analysis. 

In the upper panels of Fig.\ \ref{fig:scaling}, the asymptotic scaling behavior of $\langle \ff S_{1}\cdot \ff S_{2} \rangle$ is analyzed for the weak-$J$ limit. For the off- and the on-resonance case we find
\begin{eqnarray}
&&
\langle \ff S_{1} \cdot\ff S_{2} \rangle
\overset{J\rightarrow 0}{=} -\frac{3}{4} + \left(\frac{J}{\widetilde{J}}\right)^2 ~\text{(off resonance)} 
\; ,
\nonumber \\
&& 
\langle \ff S_{1} \cdot \ff S_{2} \rangle
\overset{J\rightarrow 0}{=} -\left(\frac{L}{L_0}\right)\times\left(\frac{J}{t}\right)  ~\text{(on resonance)} 
\; ,
\nonumber \\
\label{eq:s1s2_asymptotes}
\end{eqnarray}
respectively. 
Here, $\widetilde{J}=J_{\infty}+a/\ln(bL)$ has a weak logarithmic dependence on system size, and $L_0\approx 300$ ($a$ and $b$ are constants).
Within the perturbative regime defined by Eqs.\ (\ref{eq:rkky}) and (\ref{eq:effh}), the spin correlations cannot exhibit a $J$ dependence (there is only a single energy scale). The corrections $\propto J^{2}$ and $\propto J$ given by Eq.\ (\ref{eq:s1s2_asymptotes}) therefore reflect the $J$-dependence of terms in higher-order perturbation theory. 

%The results in the off-resonance case are compatible with an effective model including fourth-order-in-$J$ processes. 
%The linear-in-$J$ correction in the on-resonance case is more interesting: 
%Since the ground state of the effective model (\ref{eq:effh}) is a tensor product of singlets formed by the respective impurity spin with a conduction-electron spin in a Fermi orbital, a finite inter-impurity correlation $\langle \ff S_{1}\cdot \ff S_{2} \rangle < 0$ must involve breaking at least one of those singlets. The necessary energy for this process in higher-order perturbation theory is given by Eq.\ (\ref{eq:Ts_asymptotes}), and is linear in $J$.

%--------------------------------------------------------------------------------------------------------------
\section{Conclusions}
\label{sec:con}
%--------------------------------------------------------------------------------------------------------------

The physics of multiple magnetic impurities embedded in a host metal is characterized by the competition between Kondo screening and RKKY interactions. In quantum confined nanostructures, the picture changes qualitatively and the physics is far richer.
In particular, there are parity effects, and a new regime can emerge at weak coupling dominated by finite-size effects.

For a single impurity, the finite-size effects\cite{TKvD99,SA02,CB02,LVK05,HKM06} can be viewed in two complementary ways. If the conduction-electron spectrum has a gap $\Delta$ around the Fermi energy that is larger than the Kondo temperature $T_K$, then spin-flip scattering and the Kondo effect is suppressed. If there is no conduction-electron state at the Fermi energy $\varepsilon_F$ (the so-called off-resonance case), the impurity remains unscreened down to $T=0$. But in the on-resonance case, a state precisely at $\varepsilon_F$ can bind with the impurity and screen its spin through singlet formation. This is the finite-size Kondo effect, and scales linearly in the impurity-host coupling $J$. Since the finite-size gap cuts off the Kondo correlations on the scale of $\Delta$, the physics on the lowest energy scales is perturbatively accessible. 

Alternatively, one can think of the problem in real-space: if the Kondo screening cloud of extension $\xi_K$ cannot fit inside a nanostructure of finite size $L$, then the Kondo effect is replaced by its finite-size variant, in which impurity screening involves a delocalized conduction-electron state spanning the \emph{entire} system.

In finite systems with several impurities,\cite{CTL+04,ZWL+10,NBK+11} the RKKY scale must also be considered. In the off-resonance case, the impurities become coupled by the RKKY interaction on the lowest energy scales. For two impurities, they can collectively either form a singlet or a triplet state, depending on whether the RKKY interaction is antiferromagnetic or ferromagnetic. Which is realized in practice depends on details of the nanostructure geometry and electronic structure, and is encompassed by standard RKKY theory.

But, since the RKKY interaction scales as $J^2$, the finite-size Kondo effect dominates at weak coupling in the on-resonance case. This physical picture implies that for a particular system, on decreasing the coupling $J$, one can realize two successive crossovers: first from a standard Kondo strong-coupling state to the RKKY-dominated regime, and then on further decreasing $J$, to the finite-size Kondo regime. At weak coupling $J\rightarrow 0$, the finite-size Kondo effect always wins out in the on-resonance case. But if several impurities are present, can they all be screened by finite-size Kondo effects? If not, does underscreening lead to degenerate ground states? Is frustrated overscreening a possibility?

In this paper we answer these questions by putting the general concepts on firm footing. We focused mainly on the on-resonance case, deriving first an effective generalized central-spin model at weak coupling. All of the impurities (despite being located at different sites in real-space) couple to the same small number of conduction-electron states which lie precisely at $\varepsilon_F$. Several scenarios can arise, depending on the nanostructure geometry and electronic structure. We show that it is entirely possible for each impurity to be coupled to its own Fermi orbital, allowing exact finite-size-Kondo screening of all impurities, and an overall spin-singlet ground state. In other situations, there are insufficient degrees of freedom at $\varepsilon_F$ to screen all of the impurities. The remaining unscreened impurities are then coupled together via their mutual RKKY interaction. Depending on details, this could either lead to a degenerate ground state or a singlet ground state, with both finite-size-Kondo and RKKY mechanisms acting simultaneously on different impurities. We show that overscreening can never occur, since each impurity is coupled to at most one conduction-electron state at $\varepsilon_F$.

The implications of the effective low-energy effective model were substantiated by full DMRG and NRG calculations for a system comprising two impurities on a finite one-dimensional ring of conduction-electron sites. This model, although simplified, captures a number of features relevant to magnetic nanostructures. In particular, depending on the system size, one can realize either on- or off-resonance cases. Furthermore, the coupling between the impurities and the Fermi orbitals in the on-resonance case can realize either exactly screened or underscreened scenarios, depending on impurity separation.

We also investigated numerically the full crossovers between finite-size-Kondo, RKKY, and regular Kondo strong-coupling physics on increasing $J$ for a system of given size $L$, going beyond the perturbative analysis. For smaller systems, where DMRG can be used, the RKKY regime is never fully realized, since there is competition between RKKY and finite-size-Kondo effects. But for larger systems, accessible with NRG, the two crossovers are cleanly observed, with a distinct intermediate RKKY regime. The final crossover to finite-size-Kondo is pushed to weaker and weaker coupling as $L$ increases. Approaching the thermodynamic limit $L\rightarrow \infty$, this crossover is pushed to $J\rightarrow 0$, as intuitively expected.

It would be interesting to explore the physics of more realistic finite three-dimensional systems with several impurities. Analysis of the low-energy effective theory in each case should elucidate the screening mechanisms. Full numerical calculations could be implemented, using a block-Lanczos method\cite{BMF13,allerdt2015kondo,SY14} to map the real-space system onto a chain (or ladder) form amenable to treatment with either DMRG or NRG. Indeed, flatband models, \cite{Mie91a,Tas92,Tas97,STP13,TSP14} 
which sustain a macroscopically large number of states at the Fermi energy, could also be investigated.

Finally, an interesting open question regards the fate of the impurities when the conduction-electron states at $\varepsilon_F$ are not perfectly degenerate, as would arise due to symmetry-breaking perturbations (e.g.\ real-space distortions or coupling anisotropies). Then one would expect different screening mechanisms to be available at different energy scales. 
In this context, also the broadening of the energy levels due to a residual coupling of the nanostructure to the environment must be considered. \cite{SA02,LA03,Sim05,LCC12}
Another related question concerns particle-hole symmetry breaking, and the relation to hard-gapped problems,\cite{chen1998kondo} where quantum-phase transitions between Kondo and unscreened states might also be accessible.

%--------------------------------------------------------------------------------------------------------------
\acknowledgments
%--------------------------------------------------------------------------------------------------------------

Support of this work through the Deutsche Forschungsgemeinschaft within the SFB 668 (project A14)
is gratefully acknowledged.
AKM acknowledges funding from the D-ITP consortium, a program of the Netherlands Organisation for Scientific Research (NWO) that is funded by the Dutch Ministry of Education, Culture and Science (OCW). 
We are grateful for use of HPC resources at the University of Cologne.

%--------------------------------------------------------------------------------------------------------------
\appendix
%--------------------------------------------------------------------------------------------------------------

%--------------------------------------------------------------------------------------------------------------
\section{NRG calculations}
\label{sec:app_nrg}
%--------------------------------------------------------------------------------------------------------------

\begin{figure}[t]
\begin{center}
\includegraphics[width=0.8\columnwidth]{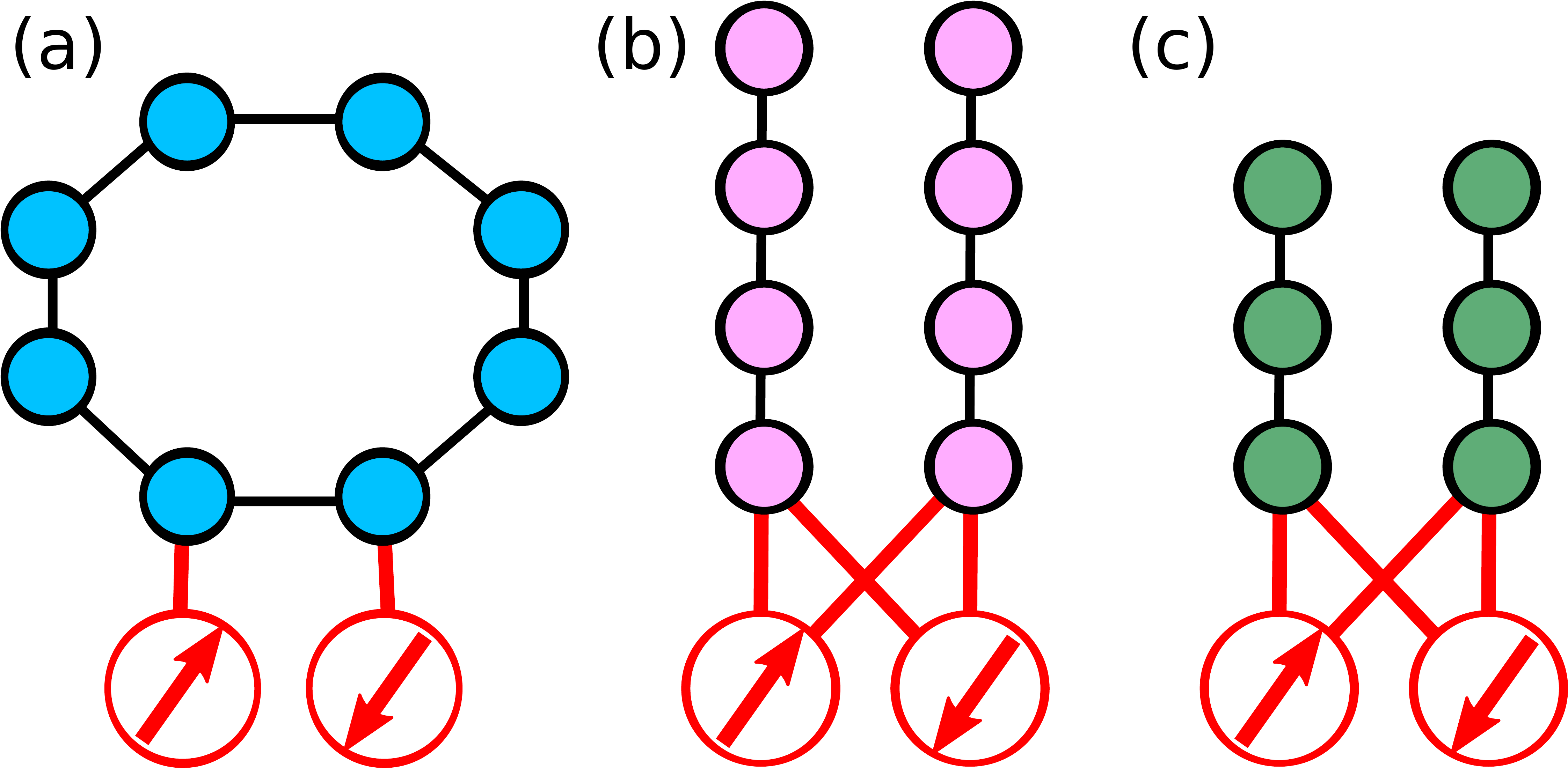}
\caption{\label{fig:schematic}
(Color online) 
Mappings for the two-impurity model in NRG. (a) Real-space basis Eq.\ (\ref{eq:H0}), with two impurities coupled to neighboring sites of a tight-binding ring comprising a finite even number of fermionic sites $L=2m$. (b) Even/odd basis Eq.\ (\ref{eq:H0_eo}), with impurities coupled to both even and odd conduction-electron channels, each comprising $L/2$ sites. (c) Discretized model Eq.\ (\ref{eq:wc}), with impurities coupled to even and odd Wilson chains, each comprising $N<L/2$ sites.
}
\end{center}
\end{figure}

In this Appendix, we describe our implementation of Wilson's NRG,\cite{Wil75,nrg_rev} generalized to treat a system comprising two spin-$\nicefrac{1}{2}$ impurities coupled to neighboring sites of a fermionic tight-binding ring of $L=2m$ sites. The physical setup is illustrated in Fig.\ \ref{fig:schematic}(a).

The conduction-electron Hamiltonian, $H_0$ in Eq.\ (\ref{eq:ham}), can be written in the form,
\begin{equation}
\label{eq:H0}
\begin{split}
H_0= &t\sum_{n=1}^{m-1}\sum_{\sigma, \alpha=\pm}  \left ( c_{\alpha n \sigma}^{\dagger}c_{\alpha (n+1) \sigma}^{\phantom{\dagger}}+\text{H.c.} \right ) \\ +& t\sum_{\sigma}\left ( c_{+1 \sigma}^{\dagger}c_{-1 \sigma}^{\phantom{\dagger}} + c_{+m \sigma}^{\dagger}c_{-m \sigma}^{\phantom{\dagger}} + \text{H.c.}\right ) \;.
\end{split}
\end{equation}
The impurity part of the Hamiltonian is $H_1=J(\textbf{S}_{+1}\cdot \textbf{s}_{+1}+\textbf{S}_{-1}\cdot \textbf{s}_{-1})$. We now transform to an even/odd orbital basis for the conduction electrons, defined in terms of the canonical fermions, $c_{(e/o)n\sigma}=\tfrac{1}{\sqrt{2}}(c_{+n\sigma}\pm c_{-n\sigma})$, viz:
\begin{equation}
\label{eq:H0_eo}
\begin{split}
H_0= &t\sum_{n=1}^{m-1}\sum_{\sigma, \alpha=e/o}  \left ( c_{\alpha n \sigma}^{\dagger}c_{\alpha (n+1) \sigma}^{\phantom{\dagger}}+\text{H.c.} \right ) \\ + &t\sum_{\sigma}\left ( c_{e 1 \sigma}^{\dagger}c_{ e 1 \sigma}^{\phantom{\dagger}}- c_{o 1 \sigma}^{\dagger}c_{ o 1 \sigma}^{\phantom{\dagger}} + c_{e m \sigma}^{\dagger}c_{e m \sigma}^{\phantom{\dagger}} - c_{o m \sigma}^{\dagger}c_{o m \sigma}^{\phantom{\dagger}} \right )
\: .
\end{split}
\end{equation}
For $J=0$, the even and odd channels are strictly decoupled. For finite $J$, each impurity couples to both even and odd channels --- see Fig.~\ref{fig:schematic}(b). This two-impurity problem is thus manifestly and irreducibly two-channel in nature. To use NRG, we must map each channel $\alpha=e/o$ into the form of a Wilson chain.\cite{Wil75}

NRG is usually used for gapless Fermi systems, and involves a logarithmic discretization of the (continuous) conduction-electron density. In the present context of the finite ring, the $J=0$ conduction-electron density of states $\rho_{\alpha}(\omega)$ for each channel $\alpha=e/o$ is already discrete,
\begin{equation}
\label{eq:LDOS}
\rho_{\alpha}(\omega) =  \sum_{p=1}^{L} a_{\alpha}(p) \delta\left [ \omega-\varepsilon(p) \right ] \;.
\end{equation}
However, the poles located at $\varepsilon(p)$ are not of course distributed on a logarithmic energy grid, as required for NRG. Employing the diagonal representation Eqs.\ (\ref{eq:umatrix}), (\ref{eq:fourier}), (\ref{eq:dispersion}) (together with the even/odd basis transformation), we find $\varepsilon(p)= 2t\cos [2\pi \tfrac{p}{L}]$ and $a_{e/o}(p)=\tfrac{1}{L}(1\pm \cos [2\pi \tfrac{p}{L}])$.

%%%%%%%%%%%%%%%%%%%%%%%%%%%%%%%%%%%%
%%%%%%%%%%%%%%%%%%%%%%%%%%%%%%%%%%%%

To perform the Wilson chain mappings, we now re-discretize (or re-bin) each spectrum $\rho_{\alpha}(\omega)$ on a logarithmic frequency grid. This is done by dividing $\rho_{\alpha}(\omega)$ into intervals of exponentially-decreasing width, defined by the discretization points $x^{\pm}_n=\pm 2t\Lambda^{-n}$ (with $n=0,1,2,3...$ and $\Lambda>1$). All poles in a given interval are replaced by a single pole of the same total weight to yield,
\begin{equation}
\label{eq:LDOS_disc}
\rho^{\text{disc}}_{\alpha}(\omega) = \sum_{n,\pm} b_{\alpha}(n,\pm) \delta\left [ \omega-\xi_{\alpha}(n,\pm) \right ] \;,
\end{equation}
where,
\begin{equation}
\label{eq:discpoles}
\begin{split}
b_{\alpha}(n,\pm) = &\sum_{p=1}^{L} a_{\alpha}(p) \theta[\pm \varepsilon(p) \mp x^{\pm}_{n+1} ] \theta[\pm x^{\pm}_{n} \mp \varepsilon(p) ]  \;, \\
\xi_{\alpha}(n,\pm) = &\sum_{p=1}^{L} \frac{\varepsilon(p) a_{\alpha}(p)}{b_{\alpha}(n,\pm)} \theta[\pm \varepsilon(p) \mp x^{\pm}_{n+1} ] \theta[\pm x^{\pm}_{n} \mp \varepsilon(p) ]  \;.
\end{split}
\end{equation}
For a given real-space system with $L$ orbitals, the number and distribution of poles in $\rho^{\text{disc}}_{\alpha}(\omega)$ is thus controlled by the discretization parameter $\Lambda$. In general there are (far) fewer poles in $\rho^{\text{disc}}_{\alpha}(\omega)$ than in the true $\rho_{\alpha}(\omega)$.

For each channel $\alpha$, the Wilson chain is now defined uniquely\cite{Wil75,nrg_rev} as the one-dimensional tight-binding chain which has the same local density of states at one end as the discretized spectrum $\rho^{\text{disc}}_{\alpha}(\omega)$,
\begin{equation}
\label{eq:wc}
\begin{split}
H_0^{\text{disc}}= \sum_{\alpha, \sigma} \Big [ &\sum_{n=0}^{N} e^{\alpha}_n f_{\alpha n \sigma}^{\dagger}f_{\alpha n \sigma}^{\phantom{\dagger}} \\+ &\sum_{n=0}^{N-1} t^{\alpha}_n \left ( f_{\alpha n \sigma}^{\dagger}f_{\alpha (n+1) \sigma}^{\phantom{\dagger}}+\text{H.c.} \right ) \Big ]\;,
\end{split}
\end{equation}
where $N\equiv N(L,\Lambda)$, $e_n^{\alpha}\equiv e_n^{\alpha}(L,\Lambda)$ and $t_n^{\alpha}\equiv t_n^{\alpha}(L,\Lambda)$ depend on the original system size $L$ and the discretization parameter $\Lambda$. The number of Wilson chain orbitals $N$ in each channel is finite and typically (far) smaller than the original number of real-space orbitals $L$.  In practice, the Wilson chain mapping is achieved by Lanczos tridiagonalization,\cite{Wil75,nrg_rev} using as input the pole weights and positions from Eq.\ (\ref{eq:discpoles}). The impurity subsystem is then coupled to the end of the Wilson chains, as depicted in Fig.~\ref{fig:schematic}(c).

%--------------------------------------------------------------------------------------------------------------
\subsection{Fermi-level pole}
%--------------------------------------------------------------------------------------------------------------

%%%%%%%%%%%%%%%%%%%%%%%%%%%%%%%%%%%%
\begin{figure}[t]
\begin{center}
\includegraphics[width=1.0\columnwidth]{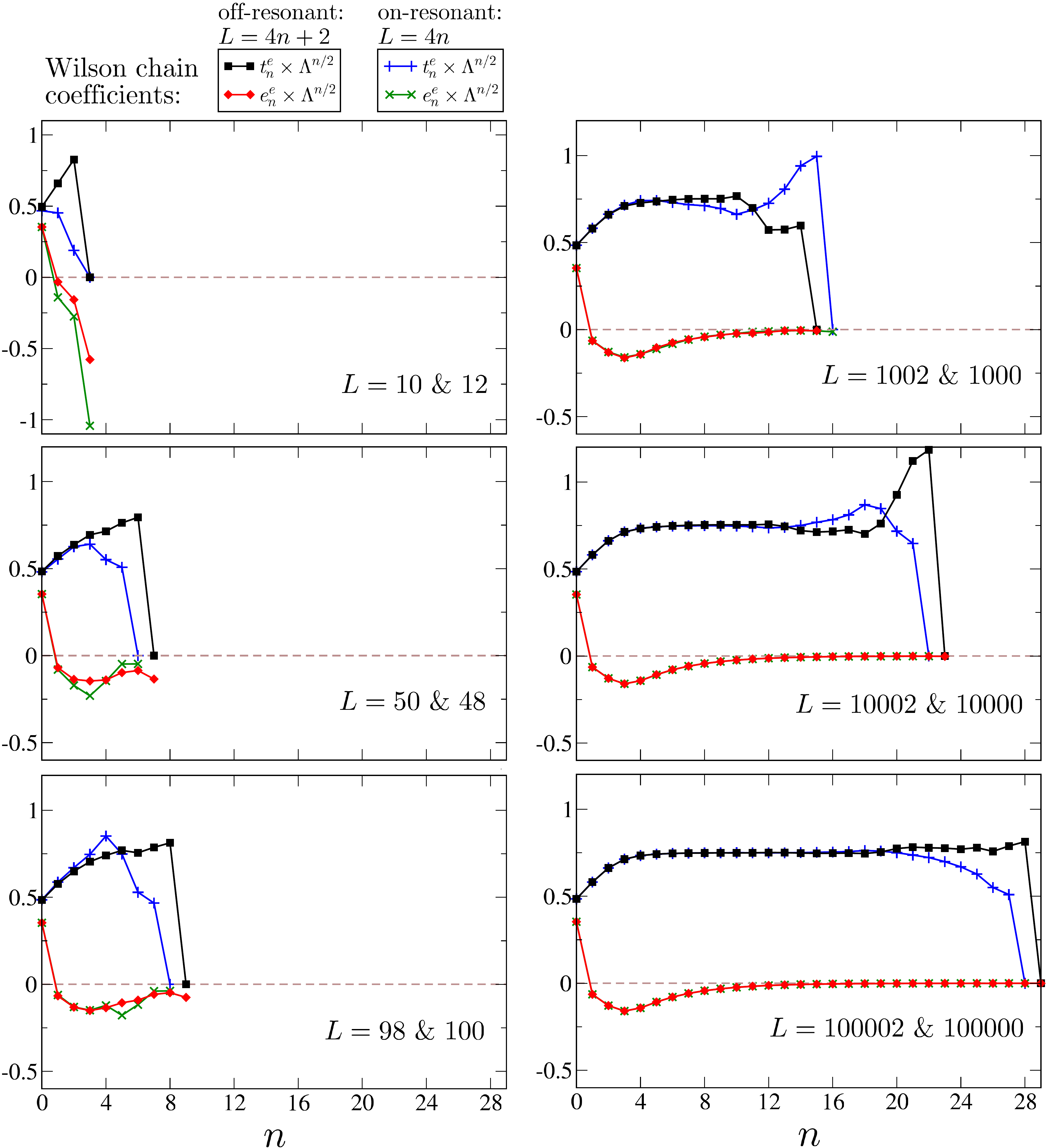}
\caption{\label{fig:wc}
(Color online) 
Wilson chain coefficients $t_n^e\times \Lambda^{n/2}$ and $e_n^e\times \Lambda^{n/2}$ vs Wilson index $n$ using $\Lambda=2$ for various system sizes $L$.
}
\end{center}
\end{figure}
%%%%%%%%%%%%%%%%%%%%%%%%%%%%%%%%%%%%

As discussed in Sec.\ \ref{sec:dmrg}, the underlying physics of the model is expected to be very different, depending on whether or not the conduction-electron system has a pole at the Fermi level. With system size $L=2m$ as above, no Fermi level pole exists in the off-resonance case for odd $m$, while a pole of weight $2/L$ lies precisely at the Fermi level in the on-resonance case for even $m$.

When there is no Fermi level pole, we can safely use the logarithmic discretization scheme embodied by Eq.\ (\ref{eq:discpoles}) to generate the Wilson chain formulation, Fig.~\ref{fig:schematic}(c).

However, in the on-resonance case, the important effect of the Fermi level pole cannot be captured by a discretization scheme defined on a \emph{logarithmic} grid. To this end, we broaden the Fermi level pole using a Lorentzian of width $\delta$. Specifically, we replace Eq.\ (\ref{eq:LDOS}) by
\begin{equation}
\label{eq:LDOSnew}
\tilde{\rho}_{\alpha}(\omega) =  \sideset{}{'}\sum_{p=1}^{L} a_{\alpha}(p) \delta\left [ \omega-\varepsilon(p) \right ] + \frac{L}{2}\times\frac{\delta /\pi}{\omega^2+\delta^2} \;,
\end{equation}
where the primed summation indicates that the pole at $\varepsilon(p)=0$ is excluded. The discretized pole positions and weights from Eq.\ (\ref{eq:discpoles}) are therefore modified to include the extra density in each interval, 
\begin{equation}
\label{eq:discpoles2}
\begin{split}
\tilde{b}_{\alpha}(n,\pm) = & b_{\alpha}(n,\pm)  \pm \frac{L}{2\pi}\arctan\left [ \frac{\delta(x^{\pm}_{n}-x^{\pm}_{n+1})}{ x^{\pm}_{n}x^{\pm}_{n+1}+\delta^2} \right ]  \\
\tilde{\xi}_{\alpha}(n,\pm)\tilde{b}_{\alpha}(n,\pm) = & \xi_{\alpha}(n,\pm) b_{\alpha}(n,\pm) \pm \frac{L \delta}{2\pi} \ln \left [ \frac{(x^{\pm}_{n})^2+\delta^2}{(x^{\pm}_{n+1})^2+\delta^2} \right ]
\end{split}
\end{equation}
Note that any finite broadening $\delta$ results in an infinite number of discretized poles (albeit with exponentially decreasing weight). The Wilson chain mapping can now be performed as before, and the results examined as a function of $\delta$.

We find that inclusion of the Fermi level pole has a significant effect on the resulting Wilson chain coefficients. However, the results converge rapidly with decreasing $\delta$. Finally, we use $\delta=10^{-10}$. Although physically this pole broadening should have negligible effect (being far smaller than any physical energy scale of the problem, including $T_K$), its inclusion does affect the Wilson chain coefficients, allowing the physics of the on-resonance case to be studied with NRG. 

Fig.~\ref{fig:wc} shows the resulting Wilson chain coefficients for the same systems as in Fig.~\ref{fig:s1s2}. Note that for both on- and off-resonance cases the Wilson chain is of finite length $N$ (meaning that $t_{N}^{\alpha}=0$). Their non-trivial evolution encodes the real-space physics of the tight-binding ring.
\\

%--------------------------------------------------------------------------------------------------------------
\subsection{Iterative diagonalization}
%--------------------------------------------------------------------------------------------------------------

The key property of the Wilson chain hoppings $t_n^{\alpha}$ is that they decay exponentially down the chain --- Fig.~\ref{fig:wc}. This is obviously not a property of the physical real-space system, but arises due to the logarithmic (re-)discretization.
As with Wilson's original NRG formulation,\cite{Wil75,nrg_rev} this justifies an iterative process of diagonalization and truncation.

Starting from the impurity subsystem, one builds up the chains by successively coupling on additional Wilson orbitals. At a given step $n\le N$, corresponding to a system comprising the impurities and $n$ Wilson orbitals of each channel, the Hamiltonian is diagonalized. However, only the lowest-energy $N_K$ states are kept for construction of the Hamiltonian at step $n+1$. Because $t_n^{\alpha}$ decay exponentially down the chain, the discarded high-energy states at iteration $n$ do not cross over into the low-energy manifold at a later iteration $n'>n$. The physics of the system is therefore examined on successively lower energy scales as more Wilson orbitals are added, embodying the RG structure of the problem.

The spin-spin correlators presented in Fig.~\ref{fig:s1s2} were obtained from the full thermal density matrix,\cite{weichselbaum2007sum} constructed in the complete Anders-Schiller basis.\cite{anders2005real}

\end{document}